\documentclass[10pt]{article}
\pdfoutput=1

\usepackage{bbm}
\usepackage[final]{graphics}
\usepackage{amsmath}
\usepackage{amsfonts,amsbsy}
\usepackage{amssymb}
\usepackage{bm}
\usepackage{color}

\def\empile#1\over#2{\mathrel{\mathop{\kern 0pt#1}\limits_{#2}}}
\def\bs{\boldsymbol}

\def\TODO#1{}

\def\p{{\boldsymbol p}}
\def\q{{\boldsymbol q}}

\def\k{{\boldsymbol k}}

\def\x{{\boldsymbol x}}
\def\y{{\boldsymbol y}}

\newcommand{\slL}{\raise.15ex\hbox{$/$}\kern-.53em\hbox{$L$}}
\newcommand{\slP}{\raise.15ex\hbox{$/$}\kern-.53em\hbox{$P$}}
\newcommand{\slD}{\raise.15ex\hbox{$/$}\kern-.67em\hbox{$D$}}
\newcommand{\slp}{\raise.1ex\hbox{$/$}\kern-.63em\hbox{$p$}}
\newcommand{\slq}{\raise.1ex\hbox{$/$}\kern-.53em\hbox{$q$}}
\newcommand{\slv}{\raise.1ex\hbox{$/$}\kern-.63em\hbox{$v$}}
\newcommand{\slR}{\raise.15ex\hbox{$/$}\kern-.53em\hbox{$R$}}
\newcommand{\slQ}{\raise.15ex\hbox{$/$}\kern-.53em\hbox{$Q$}}
\newcommand{\slK}{\raise.15ex\hbox{$/$}\kern-.53em\hbox{$K$}}
\newcommand{\slk}{\raise.15ex\hbox{$/$}\kern-.53em\hbox{$k$}}
\newcommand{\slSigma}{\raise.15ex\hbox{$/$}\kern-.53em\hbox{$\Sigma$}}
\newcommand{\slcalP}{\raise.15ex\hbox{$/$}\kern-.63em\hbox{$\cal P$}}
\newcommand{\slcalA}{\raise.15ex\hbox{$/$}\kern-.63em\hbox{$\cal A$}}
\newcommand{\slA}{\raise.15ex\hbox{$/$}\kern-.73em\hbox{$A$}}
\newcommand{\slbfA}{\raise.15ex\hbox{$/$}\kern-.73em\hbox{${\imb A}$}}
\newcommand{\slpartial}{\raise.15ex\hbox{$/$}\kern-.53em\hbox{$\partial$}}
\newcommand{\sla}{\raise.15ex\hbox{$/$}\kern-.53em\hbox{$a$}}
\newcommand{\slb}{\raise.15ex\hbox{$/$}\kern-.53em\hbox{$b$}}
\newcommand{\slc}{\raise.15ex\hbox{$/$}\kern-.53em\hbox{$c$}}
\newcommand{\slC}{\raise.15ex\hbox{$/$}\kern-.63em\hbox{$C$}}
\newcommand{\sln}{\raise.15ex\hbox{$/$}\kern-.575em\hbox{$n$}}

\newcommand\ontop[2]{\genfrac{}{}{0pt}{}{#1}{#2}}
\newcommand{\unit}{\mbox{1}\hspace{-0.25em}\mbox{l}}

\begin{document}

\title{\bf Quark~production~in~heavy~ion~collisions:\\
Formalism and boost invariant\\fermionic light-cone mode functions}
\author{Fran\c cois Gelis${}^{~a}$, Naoto Tanji${}^{~b}$}
\maketitle
 \begin{center}
   \begin{itemize}
  \item[{\bf a.}] Institut de physique th\'eorique, Universit\'e Paris Saclay\\
 CEA, CNRS, F-91191 Gif-sur-Yvette, France
  \item[{\bf b.}] Institut f\"{u}r Theoretische Physik, Universit\"{a}t Heidelberg\\ Philosophenweg 16,
    D-69120, Heidelberg, Germany
  \end{itemize}
 \end{center}

\begin{abstract}
  We revisit the problem of quark production in high energy heavy ion
  collisions, at leading order in $\alpha_s$ in the color glass
  condensate framework. In this first paper, we setup the formalism
  and express the quark spectrum in terms of a basis of solutions of
  the Dirac equation (the mode functions).  We determine analytically
  their initial value in the Fock-Schwinger gauge on a proper time
  surface $Q_s\tau_0\ll 1$, in a basis that makes manifest the boost
  invariance properties of this problem. We also describe a
  statistical algorithm to perform the sampling of the mode functions.
\end{abstract}

\section{Introduction}
\label{sec:intro}
In heavy ion collisions at ultrarelativistic energies, such as those
performed at the RHIC or the LHC, the bulk of particle production
originates from soft gluons that carry a small fraction of the
projectile longitudinal
momentum~\cite{GriboLR1,MuellQ1,BlaizM2,BlaizG1,GelisLV2,Lappi6}. Because
of the infrared singularity present in the emission probability of
soft massless gluons, the occupation number of these gluons increases
as an inverse power of the longitudinal momentum fraction $x$,
according to the BFKL evolution equation \cite{BalitL1,KuraeLF1}. When
it reaches values of order $1/\alpha_s$, non-linear processes such as
recombinations become important and tame the growth of the occupation
number -- a phenomenon known as gluon saturation \cite{GriboLR1}.

By virtue of this large occupation number, the dynamics of these soft
gluons is essentially classical, but non-perturbative because highly
non-linear \cite{McLerV1,McLerV2}. The Color Glass Condensate (CGC)
effective theory \cite{IancuLM3,Weige2,GelisIJV1} provides an
organization principle for this regime of strong interactions, and a
calculational framework for computing observables relevant to hadronic
or nuclear collisions involving such densely occupied projectiles.

In the study of heavy ion collisions, the CGC has been applied to
calculate the gluon yield at early times \cite{KrasnV2,Lappi1}. At
leading order (tree level), this amounts to solving the classical
Yang-Mills equations with light-cone currents representing the fast
color charges of the two projectiles. At next-to-leading order (one
loop) \cite{GelisLV3}, one can extract the terms that contain
logarithms of the collision energy and show that they can be absorbed
into the renormalization group evolution --according to the JIMWLK
equation \cite{IancuLM1,IancuLM2} -- of the probability distribution
of the above currents.

The CGC framework can also be used in order to study the production of
quarks in heavy ion collisions. In this framework, the light-cone
color currents couple only to gluons (because gluons are the dominant
constituents of high energy hadrons or nuclei), and quarks are
produced indirectly from the gluons by the process $\text{\sl
  gluons}\to q\overline{q}$. Thus, the quark spectrum is one order
higher in $\alpha_s$ than the gluon spectrum. Equivalently, one may
say that the quark spectrum is a 1-loop quantity while the gluon
spectrum is a tree level quantity. It is well known that the single
inclusive quark spectrum can be expressed in terms of a basis of
solutions of the Dirac equation, with a color background field that
corresponds to the LO gluons (i.e. the classical solution of the
Yang-Mills equations). The choice of this basis of solutions is not
unique. When they are chosen in such a way that they coincide with the
free spinors $v_s(\k)e^{ik\cdot x}$ or $u_s(\k)e^{-ik\cdot x}$ in the
remote past, they are often called \emph{mode functions} in the
literature, and we will also adopt this terminology in the rest of
this paper.

This approach can be used to study the production of fermions or
scalar particles in any situation where the production is due to a
classical background field
\cite{FukusGL1,Tanji1,Tanji2,Tanji3,GelisT2}.  In the context of heavy
ion collisions, this formulation\footnote{In the case of
  proton-nucleus collisions, or in any situation where it is
  legitimate to expand in powers of the color sources of one of the
  projectiles, a more direct approach is possible, that leads to
  analytical results at leading order
  \cite{GelisV1,BlaizGV1,BlaizGV2,KharzT1,KharzT4}.} has been used
first in studies of electron production in nuclear collisions
\cite{BaltzM1,BaltzGMP1} (although this is a pure QED process, its
treatment in the ``equivalent photon'' approximation is very similar
to the CGC), and later in a computation of quark production in heavy
ion collisions \cite{GelisKL1,GelisKL2}. This earlier work was limited
in a number of ways: (i) the basis of mode functions that was used was
expressed in terms of proper time and the Cartesian coordinates
$x,y,z$, making the boost invariance of the problem highly
non-obvious, (ii) the sum over the modes was restricted to a subset of
all the possible modes, and (iii) the resulting quark spectrum may be
contaminated by spurious lattice doublers at high momentum.

The goal of this work is to revisit this study in order to overcome
all these limitations. In this first paper, we first obtain a new
basis for the Dirac mode functions, that naturally depend on the
proper time $\tau$, on the rapidity $\eta$ and on the transverse
position $\x_\perp$. These mode functions are indexed by the
transverse momentum $\k_\perp$ and a wave number $\nu$ which is the
Fourier conjugate to $\eta$, making them very convenient for a lattice
implementation where the grid covers a fixed range in $\eta$. In order
to improve the sampling of the mode functions, we use spinors that are
random linear superpositions of all the possible mode functions. A
proper choice of the distribution of the random weights ensures that
the exact result is recovered in the limit of infinite
statistics. With finite statistics, this procedure provides a
straightforward way to estimate the statistical errors. Numerical
results based on a lattice implementation of this framework will be
presented in a forthcoming paper.

The contents of the paper is the following~: In the section
\ref{sec:cgc}, we briefly remind the reader of the Color Glass
Condensate and of the expression of the quark spectrum in this
framework. We also show in this section how to choose a basis of mode
function that makes boost invariance manifest. In the section
\ref{sec:stat}, we present a statistical method to sample the modes,
and derive the formula for the corresponding statistical errors. The
initial value of the mode functions on the forward light-cone
(i.e. just after the collision of the two nuclei) is derived in the
section \ref{sec:mode}. The section \ref{sec:summary} is devoted to
concluding remarks. A few appendices collect more technical
material. The derivation of the expression of the quark inclusive
spectrum in terms of Dirac mode functions is recalled in the appendix
\ref{sec:spectrum}, and an alternate derivation following more closely
standard Feynman perturbation theory is presented in the appendix
\ref{sec:spectrum-F}. The appendix \ref{app:IC} discusses a
technicality in the derivation of the initial value of the mode
functions, and the appendix \ref{app:inner} is devoted to the study of
a conserved inner product between the mode functions. We make an
extensive use of this inner product in order to properly normalize the
mode functions, and as a consistency check at various stages of the
calculation.  In the section \ref{sec:abelian}, we use the QED version
of the mode functions derived in the section \ref{sec:mode} in order
to recover the electron production amplitude in the collision of two
electrical charges.

\section{Quark yield in the CGC framework}
\label{sec:cgc}
\subsection{Color Glass Condensate}
The Color Glass Condensate framework is an effective theory that can
be used to study the early stages of heavy ion collisions, summarized
by the following Lagrangian density,
\begin{equation}
{\cal L}=-\frac{1}{4}F^{\mu\nu}F_{\mu\nu} + A_\mu(J_1^\mu+J_2^\mu)+\overline{\psi}(i\slD-m)\psi\; ,
\end{equation}
written here for one family\footnote{As long as we do not include the
  effect of virtual quark loops, we can consider one quark family at a
  time.} of quarks of mass $m$. The color charge content of the
incoming nuclei is described by the two currents $J_{1,2}^\mu$,
  whose supports are restricted to the light-cones,
  $J_1^\mu\propto\delta(x^-)$ and $J_2^\mu\propto \delta(x^+)$, in a
  collision at very high energy. These currents fluctuate
event-by-event, with a Gaussian probability distribution in the
McLerran-Venugopalan (MV) model\footnote{At very high energies, this
  distribution evolves according to the JIMWLK equation and will
  become non-Gaussian. The MV model may be viewed as a model of
  initial condition for the JIMWLK evolution.}  \cite{McLerV1,McLerV2}
that we use in this paper. If one is interested in the production of
quarks in a given collision, one could draw randomly one configuration
of $J_{1,2}^\mu$, and not perform an average over these
currents\footnote{Note however that the theoretical basis for doing
  this is not very robust, and becomes inconsistent beyond leading
  order. Indeed, for fixed $J_{1,2}^\mu$, loop corrections to
  observables contain unphysical logarithms of the longitudinal
  momentum cutoff that separate the gluon modes that are described by
  the sources $J^\mu$ and those that are described as gauge
  fields. These logarithms can be absorbed into a redefinition of the
  probability distribution $W[J]$ of these sources (that must now
  evolve according to the JIMWLK equation). But this procedure only
  works if one performs an average over the sources.}.

In the gluon saturation regime, the currents $J_{1,2}$ are inversely
proportional to the gauge coupling,
\begin{equation}
J_{1,2}^\mu \sim \frac{1}{g}\; .
\end{equation}
For this reason, gluonic observables at leading order are expressible
in terms of a classical color field that obeys the Yang-Mills equation
with the source $J_1+J_2$,
\begin{equation}
\big[D_\mu ,F^{\mu\nu}\big]=J_1^\mu+J_2^\mu\; .
\label{eq:YM}
\end{equation}
One should in principle also impose the covariant conservation of the
current $[D_\mu,J_1^\mu+J_2^\mu]=0$. This constraint becomes trivial
in the Fock-Schwinger gauge, $x^+A^-+x^-A^+=0$, since it ensures that
the gauge potential vanishes on the support of the current. We adopt
this gauge in the following.  Furthermore, for inclusive observables,
one can prove that this equation of motion must be supplemented by a
retarded boundary condition \cite{GelisV2}, such that the gauge field
vanishes in the remote past, thereby making this classical solution
unique. In the saturation regime, this classical gauge field is
strong, of order $A^\mu\sim 1/g$.

\subsection{Inclusive quark spectrum}
In the CGC framework, the fermions do not couple directly to the
currents $J_{1,2}^\mu$, but only indirectly through the gauge field
that appears in the covariant derivative in the Dirac operator
$i\slD-m$. Therefore, the natural order of magnitude of the spinors is
$\psi\sim 1$, in accordance with the fact that the occupation number
of fermions is bounded by unity. Thus, observables that contain quark
fields are of higher order in the gauge coupling. For instance, the
$g^2$ power counting for the quark spectrum at LO is the same as that
of the gluon spectrum at NLO: both are 1-loop quantities, the only
difference being the nature of the field running in the loop (quark
versus gluon). In a fixed background color field, the quark spectrum
at LO is given by the following formula\footnote{This formula is true
  to all orders in the currents $J_1^\mu$ and $J_2^\mu$. If one
  expands it to lowest order in these currents (i.e. $dN_{\rm
    q}/d^3\p\propto {\cal O}((J_1J_2)^2)$), one recovers the standard
  result for the process $gg\to q\overline{q}$ with off-shell incoming
  gluons, derived in the framework of $k_T$-factorization in
  refs.~\cite{CatanCH1,ColliE1} (see ref.~\cite{GelisV1} for this
  comparison).}, whose derivation is recalled in the appendix
\ref{sec:spectrum}~:
\begin{eqnarray}
2\omega_\p\frac{dN_{\rm q}}{d^3\p}
&=&
\frac{1}{(2\pi)^3}
\sum_{\ontop{\sigma,s=\uparrow,\downarrow}{a,b}}\int
\frac{d^3\k}{(2\pi)^3 2\omega_\k}\;
\lim_{x^0\to +\infty}\left|
\big(\psi^{0+}_{\p\sigma b}\big|\psi^-_{\k s a}\big)_{x^0}
\right|^2\; ,
\label{eq:Q-LO0}
\end{eqnarray}
($\omega_\k\equiv\sqrt{\k^2+m^2}$) where $\psi_{\p \sigma b}^{0+}$ is
a free positive energy spinor of momentum $\p$, spin $\sigma$ and
color $b$ (since quarks live in the fundamental representation of the
gauge group $SU(N_c)$, this color index runs from 1 to $N_c$). In the
absence of background field, these spinors are given by\footnote{In
  this equation and in the rest of this paper, we write explicitly
  only the indices that characterize the initial value of the
  spinor. A more complete notation would read~:
\begin{equation*}
\psi_{\p \sigma b}^{0+\alpha b'}(x) = u_\sigma^\alpha(\p)\,\delta_{bb'}\,e^{-ip\cdot x}\; ,
\end{equation*}
where $\alpha$ is the Dirac index and $b'$ is the color of the quark
at the point $x$.  },
\begin{equation}
\psi_{\p \sigma b}^{0+}(x) = u_\sigma(\p)\,e^{-ip\cdot x}\qquad(p^0=\omega_\p)\; .
  \label{eq:psi00}
\end{equation}
However, it may also happen that the gauge fields at $x^0\to+\infty$
evolve into a nonzero pure gauge configuration. In this case, the
above spinor should be replaced by a color rotated one:
\begin{equation}
  \psi_{\p \sigma b}^{0+\alpha b'}(x) = u_\sigma^\alpha(\p)\,\Omega_{b'b''}(\x)\delta_{bb''}\,e^{-ip\cdot x}\; ,
  \label{eq:psi0Omega}
\end{equation}
where $SU(N_c)$ is the $SU(N_c)$ matrix defining the pure gauge background.

In contrast, $\psi_{\k s a}^-$ is a spinor that has evolved over the
background color field, starting at $x^0=-\infty$ from a negative
energy free spinor of momentum $\k$ and spin $s$~:
\begin{equation}
(i\slD_x-m)\,\psi^-_{\k s a}(x) = 0\quad,\quad\lim_{x^0\to -\infty}\psi^-_{\k s a}(x)
= v_s(\k)e^{ik\cdot x}\; .
\end{equation}
Note that the subscripts $a,b$ refer to the initial color of the
quarks. The color they carry at the point $x$ is not written
explicitly, and is encoded in the $N_c\, (\mbox{color})\times 4\,
(\mbox{Dirac})$ components of the spinors.

The inner product $\big(\cdot\big|\cdot\big)_{x^0}$ that appears under
the integral in eq.~(\ref{eq:Q-LO0}) is defined by
\begin{equation}
\big(\psi\big|\chi\big)_{x^0}
\equiv \int d^3\x\; \psi^\dagger(x^0,\x)\,\chi(x^0,\x)\; .
\end{equation}
(In the product $\psi^\dagger\chi$, all the unwritten color and Dirac
indices are contracted.) The properties of this inner product are
studied in detail in the appendix \ref{app:inner}. In this appendix,
we also use its conservation as a consistency check of the results
that will be derived in the section \ref{sec:mode}. Note that the
formula for the quark spectrum requires that one takes the limit of
infinite time. As we shall discuss later in this section, this is also
a requirement for the quark spectrum to be gauge invariant.

Eq.~(\ref{eq:Q-LO0}) is the expression for the fully inclusive
spectrum that we are going to use in the rest of this paper.  The
virtue of this formula is that it reduces the calculation of a
one-loop graph in a background field to solving a (linear) partial
differential equation with retarded boundary conditions. Even if this
can be done analytically only for very simple backgrounds, this
problem can in principle be tackled numerically for completely general
backgrounds.

Note that in eq.~(\ref{eq:Q-LO0}), the spectrum is summed over all the
possible final states and over the spin of the tagged quark.  In order
to obtain the polarized spectrum, one simply needs to remove the sum
over the spin index $\sigma$. This formula also contains sums over the
colors of the initial and final fermion. These sums should not be
undone, as the spectrum of quarks with a given color has no gauge
invariant meaning. $\k$ and $s$ can be viewed as the momentum and spin
of the antiquark that must be produced along with the quark to satisfy
the conservation of the flavor quantum number, as reflected by the
initial condition for the spinor $\psi_{\k s}$ in the remote past.

\subsection{Boost invariance}
\label{sec:boost}
\subsubsection{Change of coordinates}
Since collisions in the high energy limit are invariant under boosts
along the longitudinal axis\footnote{Here, we are disregarding the
  small-$x$ evolution of the color sources in the incoming
  nuclei. This effect would break the boost invariance of the problem
  due to gluon loop corrections, and make the quark spectrum depend on
  rapidity on scales $\Delta y\sim \alpha_s^{-1}$.}, it is convenient
to trade the longitudinal components of the momenta $p_z, k_z$ in
favor of the corresponding rapidities $y_p$ and $y_k$. Likewise, the
proper time $\tau$ and spatial rapidity $\eta$ are more suitable than
$x^0$ and $z$ to map the space-time:
\begin{equation}
\tau\equiv\sqrt{t^2-z^2}\quad,\quad \eta\equiv \frac{1}{2}\ln\left(\frac{t+z}{t-z}\right)\; .
\end{equation}
Besides the obvious change in the measure $dp_z/\omega_\p = dy$, one
must alter the definition of the inner product so that the integration
is on a surface of constant $\tau$ (instead of a constant $x^0$),
\begin{equation}
\big(\psi\big|\chi\big)_\tau\equiv \tau\int d^2\x_\perp d\eta\; \psi^\dagger(\tau,\x_\perp,\eta)\,e^{-\eta\gamma^0\gamma^3}\,\chi(\tau,\x_\perp,\eta)\; .
\label{eq:inner}
\end{equation}
(See the appendix \ref{app:inner}.)  When doing this,
eq.~(\ref{eq:Q-LO0}) becomes
\begin{equation}
\frac{dN_{\rm q}}{dy_p d^2\p_\perp}
=
\frac{1}{8\pi(2\pi)^3}
\sum_{\ontop{\sigma,s=\uparrow,\downarrow}{a,b}}\int
\frac{d^2\k_\perp}{(2\pi)^2}dy_k\;
\lim_{\tau\to +\infty}
\left|\big(
\psi^{0+}_{\p_\perp y_p \sigma b}
\big|
\psi^-_{\k_\perp y_k s a}
\big)_\tau\right|^2\; .
\label{eq:Q-LO2}
\end{equation}
The boost invariance of the problem implies that the inner product
depends only on the difference of the rapidities $y_p-y_k$. After
integration over $y_k$, the resulting quark spectrum is independent of
the rapidity $y_p$.

\subsubsection{Boost invariant spinors}
The boost invariance can be made manifest at the level of the spinors
$\psi_{\p_\perp y_p \sigma b}^{0+}$ and $\psi_{\k_\perp y_k s a}^-$
themselves. Even when the background field is invariant under
  boosts in the $z$ direction, these spinors depend separately on the
  momentum rapidity $y$ and on the spacetime rapidity $\eta$. This can
  be trivially seen on the vacuum spinors, whose rapidity dependence
  can be made explicit as follows
\begin{equation}
\psi_{\p_\perp y_p \sigma b}^{0+}(\tau,\eta,\x_\perp)
=
e^{\tfrac{y_p}{2}\gamma^0\gamma^3}\;
u_\sigma(\p_\perp,y=0)\;e^{-i M_\p\tau\cosh(y_p-\eta)}\,e^{i\p_\perp\cdot\x_\perp}\; ,
\end{equation}
where $M_\p\equiv\sqrt{\p_\perp^2+m^2}$ is the transverse mass. To
turn the prefactor into a function of $y_p-\eta$, it is convenient to
define transformed spinors as follows,
\begin{equation}
\widehat\psi{}_{\k_\perp y s a} \equiv \sqrt{\tau}\,e^{-\frac{\eta}{2}\gamma^0\gamma^3}\;\psi{}_{\k_\perp y s a}\; .
\label{eq:BI-spinors}
\end{equation}
The factor $\sqrt{\tau}$ has been introduced for later
convenience. After this transformation, the new spinors
$\widehat\psi_{\k_\perp y s a}$ are boost invariant, in the sense that
they depend on the spatial rapidity $\eta$ and on the momentum
rapidity $y$ only through the difference $y-\eta$ (provided that the
background field does not depend on $\eta$).

The boosted spinors introduced in eq.~(\ref{eq:BI-spinors}) also
offer the advantage of obeying a simpler form of the Dirac equation
where rapidity does not appear explicitly in the coefficients. In
order to see this, first note that
\begin{equation}
\gamma^0\partial_0+\gamma^3\partial_3
=
\gamma^0\;e^{-\eta\gamma^0\gamma^3}\;\partial_\tau
+\frac{1}{\tau}\gamma^3\;e^{-\eta\gamma^0\gamma^3}\;\partial_\eta\; .
\end{equation}
Then, multiply this operator on the left by
$\exp(-\frac{\eta}{2}\gamma^0\gamma^3)$. A simple calculation gives
\begin{equation}
e^{-\frac{\eta}{2}\gamma^0\gamma^3}\,
\left[\gamma^0\partial_0+\gamma^3\partial_3\right]
=
\left[\gamma^0\,\partial_\tau
+\frac{\gamma^3}{\tau}\,\partial_\eta+\frac{\gamma^0}{2\tau}\right]\;
e^{-\frac{\eta}{2}\gamma^0\gamma^3}\; .
\end{equation}
From this observation, we conclude that the modified spinors
$\widehat\psi$ obey the following Dirac equation~:
\begin{equation}
\left[i\left(\gamma^0\,D_\tau
+\frac{\gamma^3}{\tau}\,D_\eta+\gamma^iD_i\right)-m\right]
\;\widehat\psi=0\; .
\label{eq:dirac-mod}
\end{equation}

One can see that the coefficients of this equation are independent of
the rapidity $\eta$ when the background field is boost invariant (so
that there is no $\eta$ dependence hidden in the covariant
derivatives).  In terms of the boost invariant spinors defined in
eq.~(\ref{eq:BI-spinors}), the inner product on a constant proper time
surface takes a particularly simple form,
\begin{equation}
\big(\psi\big|\chi\big)_\tau\equiv \int d^2\x_\perp d\eta\; \widehat\psi^\dagger(\tau,\x_\perp,\eta)\,\widehat\chi(\tau,\x_\perp,\eta)\; .
\label{eq:inner1}
\end{equation}

\subsubsection{Mode functions in the $\nu$ basis}
Another useful transformation is to go from a basis where the spinors
have a definite momentum rapidity $y$ to a basis where they have a
fixed Fourier conjugate $\nu$ to the space-time rapidity $\eta$,
\begin{equation}
\widehat\psi_{\k_\perp y s a}\to 
\widehat\psi_{\k_\perp \nu s a}\equiv\int dy\; e^{i\nu y}\; \widehat\psi_{\k_\perp y s a}\; .
\end{equation} 
When the background field is boost invariant (i.e. independent of
$\eta$), $\widehat\psi_{\k_\perp y s a}$ depends only on $y-\eta$ and
the spinor $\widehat\psi_{\k_\perp \nu s a}$ in the new basis has a
trivial $\eta$ dependence in $\exp(i\nu\eta)$:
\begin{equation}
\widehat\psi_{\k_\perp \nu s a}(\tau,\x_\perp,\eta)
=
e^{i\nu\eta}\;\widetilde\psi_{\k_\perp \nu s a}(\tau,\x_\perp)\; .
\end{equation}
$\nu$ is a conserved quantum number and the $\eta$ dependence of these
spinors is not altered by their propagation over the background
field. Moreover, the Dirac equation obeyed by these spinors is
effectively $2+1$ dimensional,
\begin{equation}
\left[i\left(\gamma^0\,D_\tau
+i\frac{\gamma^3}{\tau}\,(\nu-gA_\eta)+\gamma^iD_i\right)-m\right]
\;\widetilde\psi_{\k_\perp \nu s a}=0\; ,
\label{eq:dirac-2d}
\end{equation}
since the $\eta$ dependence can be factored out.

When calculating the inner product of two such spinors (see
eq.~(\ref{eq:inner-hatpsi})), the integration over $\eta$ trivially
yields a delta function,
\begin{equation}
\big(\psi_{\p_\perp \nu \sigma b}\big|\psi_{\k_\perp \nu' s a}\big)_\tau
=
2\pi\delta(\nu-\nu')\;
\underbrace{\int d^2x_\perp\;
\widetilde\psi^\dagger_{\p_\perp \nu \sigma b}(\tau,\x_\perp)
\widetilde\psi_{\k_\perp \nu' s a}(\tau,\x_\perp)}_{\equiv\big[\psi_{\p_\perp \nu \sigma b}\big|\psi_{\k_\perp \nu' s a}\big]_\tau}\; ,
\end{equation}
where we denote by $\big[\cdot\big|\cdot\big]_\tau$ the ``reduced''
inner product that remains after one has factored out the delta function.
In this basis, the quark spectrum is given by
\begin{equation}
\frac{dN_{\rm q}}{dy_p d^2\p_\perp}
=
\frac{1}{8\pi(2\pi)^3}
\sum_{\ontop{\sigma,s=\uparrow,\downarrow}{a,b}}\int
\frac{d^2\k_\perp}{(2\pi)^2}\frac{d\nu}{2\pi}\;
\lim_{\tau\to+\infty}\left|\big[
\psi_{\p_\perp \nu \sigma b}^{0+}
\big|
\psi_{\k_\perp \nu s a}^{-}
\big]_\tau\right|^2\; .
\label{eq:Q-LO3}
\end{equation}
In this formula, it is tempting to ignore the limit $\tau\to+\infty$
and to interpret the resulting expression as the quark spectrum at the
finite proper time $\tau$. One should however consider such a
generalization with caution, since it is not possible to rigorously
define asymptotic states at a finite time.

\subsection{Gauge invariance}
\label{subsec:gauge}
Under a local gauge transformation, a spinor $\psi(x)$ is
transformed as follows
\begin{equation}
\psi(x)\quad\to\quad \Omega(x)\,\psi(x)\; ,
\end{equation}
where $\Omega(x)$ is an $SU(N_c)$ matrix. Since the inner product that
enters in the quark spectrum given by eqs.~(\ref{eq:Q-LO0}) or
(\ref{eq:Q-LO3}) is local, it is gauge invariant provided of course
that the spinors $\psi^-$ and $\psi^{0+\dagger}$ are gauge rotated
consistently.  

In eq.~(\ref{eq:psi0Omega}), we have already indicated that the
spinors $\psi^{0+\dagger}$ should be obtained from the free solutions
in a null background (\ref{eq:psi00}) by an appropriate color
rotation, if the background field at the time of quark measurement is
a nonzero pure gauge. The square of the inner product that appears in
eq.~(\ref{eq:Q-LO0}) can be written as
\begin{eqnarray}
\left|
\big(\psi^{0+}_{\p\sigma b}\big|\psi^-_{\k s a}\big)_{x^0}
\right|^2
&=&
\int d^3\x d^3\y\;e^{i\p\cdot(\y-\x)}\;
\Big[\psi^-_{\k s a}(x^0,\y)\Big]^\dagger u_\sigma(\p)\nonumber\\
&&\qquad\qquad\times
\Big[\Omega(\y)\Omega^\dagger(\x)\Big] u_\sigma^\dagger(\p)\psi^-_{\k s a}(x^0,\x)
\; .
\label{eq:tmp00}
\end{eqnarray}
For the time being, let assume that the background field is a pure
gauge at the time $x^0$ where the quarks are being
measured\footnote{This should be the case at least when
  $x^0\to+\infty$ in a sensible model of the gauge fields produced in
  heavy ion collisions, thanks to the expansion and dilution of the
  system.}.  The factor $\Omega(\y)\Omega^\dagger(\x)$ depends on this
pure gauge, and can be obtained by a Wilson line between the
points $\x$ and $\y$,
\begin{equation}
  \Omega(\y)\Omega^\dagger(\x)
  =U_\gamma(\y,\x)
  \equiv{\rm P}\,\exp\Big( ig\int_\gamma dz^\mu A_\mu(z)\Big)\; ,
  \label{eq:WL}
\end{equation}
where $\gamma$ is a path from $\x$ to $\y$ in the hyperplane of fixed
time $x^0$. Thus, in practice one would calculate
\begin{eqnarray}
\left|
\big(\psi^{0+}_{\p\sigma b}\big|\psi^-_{\k s a}\big)_{x^0}
\right|^2
&=&
\int d^3\x d^3\y\;e^{i\p\cdot(\y-\x)}\;
\Big[\psi^-_{\k s a}(x^0,\y)\Big]^\dagger u_\sigma(\p)\nonumber\\
&&\qquad\qquad\times\;
U_\gamma(\y,\x)\; u_\sigma^\dagger(\p)\psi^-_{\k s a}(x^0,\x)
\; .
\label{eq:tmp01}
\end{eqnarray}

When the background field is a pure gauge, this does not depend
on the path $\gamma$ chosen between $\x$ and $\y$. If one extends
the use of eq.~(\ref{eq:tmp01}) to a situation where the background
field at the time $x^0$ is not a pure gauge, one still obtains a gauge
invariant result, but there is now an ambiguity due to the choice of
the path.  Indeed, Wilson lines $U_{\gamma}$ and $U_{\gamma'}$
evaluated on two different paths differ by a Wilson loop,
\begin{equation}
U_\gamma U^\dagger_{\gamma'}={\rm P}\,\exp\,\Big(
ig\oint_{\gamma\cup{\gamma'}^{-1}} dz^\mu\;A_\mu(z)\Big)
\end{equation}
defined over the closed loop $\gamma\cup{\gamma'}^{-1}$ made of the
path $\gamma$ followed by the reverse of the path $\gamma'$. This
Wilson loop measures the chromo-magnetic flux across the closed loop,
and is therefore equal to the identity only if the background field is
a pure gauge. One expects this irreducible ambiguity to decrease with
the quark momentum. On the one hand, in the spectrum of quarks of
momentum $\p$, the typical spatial separation $\big|\x-\y\big|$ is of
order $1/|\p|$ (since $\x-\y$ and $\p$ are Fourier conjugates in
eq.~(\ref{eq:tmp00})), and the closed loops that one would have to
consider in the above argument have a typical area of order
$1/|\p|^2$. On the other hand, numerical studies of the gauge fields
produced in the McLerran-Venugopalan model indicate that the
expectation value of Wilson loops decreases exponentially with the
area of the loop when it becomes larger than
$Q_s^{-2}$~\cite{DumitNP1,DumitLN1}. This suggests that this ambiguity
should not affect much the quark spectrum for $|\p|\gtrsim Q_s$.

\section{Statistical sampling method}
\label{sec:stat}
\subsection{Sketch of a direct algorithm}
Eq.~(\ref{eq:Q-LO3}) contains the essence of our procedure for
calculating the quark spectrum:
\begin{itemize}
\item[{\bf i.}] Draw randomly a pair of sources $J_{1,2}^\mu$, and
  solve the classical Yang-Mills equation (\ref{eq:YM}) with null
  retarded initial conditions.
\item[{\bf ii.}] For a given transverse momentum $\k_\perp$,
  wavenumber $\nu$, spin $s$ and color $a$, initialize the spinor
  $\psi_{\k_\perp \nu s a}$ as a free negative energy spinor.
\item[{\bf iii.}] Solve the reduced 2+1 dimensional Dirac
  equation~(\ref{eq:dirac-2d}) for the time evolution of this spinor
  over the color field found in the step {\bf i}.
\item[{\bf iv.}] At some sufficiently large time, project the spinor
  $\psi_{\k_\perp \nu s a}$ on a free positive energy spinor of
  transverse momentum $\p_\perp$, same wavenumber $\nu$, spin $\sigma$
  and color $b$. This gives the reduced inner product $\big[
  \psi_{\p_\perp \nu \sigma b}^{0+} \big| \psi_{\k_\perp \nu s a}^{-}
  \big]_\tau$.
\item[{\bf v.}] Repeat the steps {\bf ii} to {\bf iv} in order to sum
  over all $\k_\perp,\nu,s,a$'s.
\item[{\bf vi.}] (optional) Repeat steps {\bf i} to {\bf v} in order
  to average over the configurations of the color sources $J_{1,2}^\mu$.
\end{itemize}

If we discretize the transverse plane into a $N_x\times N_y$ grid and the
rapidity axis into a grid with $N_\eta$ points, the computational cost for
solving the Dirac equation for one mode function scales as $N_xN_y
N_\tau$ where $N_\tau$ is the number of timesteps. Note that this cost
does not depend on $N_\eta$ since the $\eta$ dependence can be factored out
for a boost invariant background.  This must be repeated for the
$N_xN_yN_\eta$ mode functions. Therefore, the total computational cost scales
as $(N_xN_y)^2N_\eta N_\tau$.

\subsection{Statistical  method}
\label{sec:statmethod}
The algorithm described in the previous subsection is deterministic, but
it suffers from an unfavorable scaling with the size of the transverse
grid, since the computational cost scales as $(N_xN_y)^2$. Instead of doing
in full the sum over all the modes, it is possible to do it by a
Monte-Carlo sampling in which one uses random linear superpositions of
the mode functions
\begin{equation}
\psi_{c}^- \equiv \sum_{\ontop{s=\uparrow,\downarrow}{a}} 
\int \frac{d^2\k_\perp}{(2\pi)^2}\frac{d\nu}{2\pi}\;c_{\k_\perp \nu s a}\;
\psi_{\k_\perp \nu sa}^-\; ,
\label{eq:MC1}
\end{equation}
where the coefficients $c_{\k_\perp \nu s a}$ are random numbers with
the following variance
\begin{equation}
\big<c_{\k_\perp \nu s a}c_{\k'_\perp \nu' s' a'}^*\big>= (2\pi)^3\,
\delta(\k_\perp-\k'_\perp)\,\delta(\nu-\nu')\,\delta_{ss'}\,\delta_{aa'}\; .
\label{eq:variance}
\end{equation}
The justification of this approach is to note that the projector on
the subspace of negative energy spinors can be rewritten as follows
\begin{eqnarray}
\sum_{\ontop{s=\uparrow,\downarrow}{a}}\int\frac{d^2\k_\perp}{(2\pi)^2}\frac{d\nu}{2\pi}\;
\big|\psi^-_{\k_\perp \nu s a}\big)
\big(\psi^-_{\k_\perp \nu s a}\big|
=
\int [Dc]\;{\cal P}[c]\;\;
\big|\psi^-_{c}\big)
\big(\psi^-_{c}\big|\; ,
\label{eq:stat1}
\end{eqnarray}
where ${\cal P}[c]$ is a normalized probability distribution (in
practice a Gaussian distribution) with zero mean value and a variance
given in eq.~(\ref{eq:variance}). In eq.~(\ref{eq:Q-LO3}), the
integrals over $\k_\perp,\nu$ and the sums over $s,a$ are then replaced
by a statistical average over these random numbers,
\begin{equation}
\underbrace{2\pi\delta(0)}_{L_\eta}\frac{dN_{\rm q}}{dy_p d^2\p_\perp}
=
\frac{1}{8\pi(2\pi)^3}
\sum_{\ontop{\sigma=\uparrow,\downarrow}{b}}\int\frac{d\nu}{2\pi}[Dc]\;{\cal P}[c]\;
\lim_{\tau\to+\infty}
\left|\big(
\psi_{\p_\perp \nu \sigma b}^{0+}
\big|
\psi_{c}^{-}
\big)_\tau\right|^2.
\label{eq:Q-LO4}
\end{equation}
($L_\eta$ is the total length of the $\eta$ interval represented in
the lattice implementation.)  Since the random sum in
eq.~(\ref{eq:MC1}) mixes\footnote{One may replace eq.~(\ref{eq:MC1})
  by a random sum in which the modes $\nu$ are not mixed. These
  restricted linear superpositions obey the reduced 2+1 dimensional
  Dirac equation (\ref{eq:dirac-2d}), but one must repeat the
  resolution of the equation for each mode $\nu$, so that this
  modification has no merit in terms of computational cost. In fact,
  using eq.~(\ref{eq:MC1}) and solving the 3+1 dimensional Dirac
  equation has the advantage that it is immediately generalizable to a
  non-boost invariant background color field if necessary.}  the
various $\nu$'s, the evolution of $\psi_{c}^-$ is governed by the 3+1
dimensional Dirac equation (\ref{eq:dirac-mod}), and the computational
cost of this approach scales as $N_xN_yN_\eta N_\tau N_{\rm conf}$ where
$N_{\rm conf}$ is the number of samples used in the statistical
average. Compared to the direct deterministic method, a power of $N_xN_y$
has been replaced by $N_{\rm conf}$, which is advantageous for
large grids if $N_{\rm conf}\ll N_xN_y$.

\subsection{Statistical errors}
\label{sec:staterr}
The statistical method summarized by the eqs.~(\ref{eq:stat1}) and
(\ref{eq:Q-LO4}) is exact only in the case of a perfect sampling of
the Gaussian distribution ${\cal P}[c]$. In practice, this sampling is
performed by generating a large but finite number $N_{\rm conf}$ of
configurations. In doing this, the left hand side of
eq.~(\ref{eq:stat1}) is replaced by
\begin{equation}
\frac{1}{V^2}\sum_{\vec{\bs J},\vec{\bs J}'}
\;C_{N_{\rm conf}}(\vec{\bs J},\vec{\bs J}')\;
\big|\psi^-_{\vec{\bs J}}\big)
\big(\psi^-_{\vec{\bs J}'}\big|
\label{eq:stat3}
\end{equation}
where we have used discrete notations that correspond to the lattice
implementation, and we use the following shorthands~:
\begin{eqnarray}
V&\equiv&L_x L_y L_\eta \qquad\mbox{(total lattice volume)}\nonumber\\
\vec{\bs J}&\equiv&(j_x,j_y,j_\eta,s,a)\; .
\end{eqnarray}
(The integers $j_{x,y,\eta}$ label the momentum modes $\k_\perp,\nu$
in the lattice implementation, and $s,a$ are the spin and color
quantum numbers.) The coefficients that appear in the sum in
eq.~(\ref{eq:stat3}) result from $N_{\rm conf}$ samplings of the
Gaussian distribution~:
\begin{equation}
C_{N_{\rm conf}}(\vec{\bs J},\vec{\bs J}')\equiv
\frac{1}{N_{\rm conf}}\sum_{n=1}^{N_{\rm conf}}c_{\vec{\bs J}}^{(n)}c_{\vec{\bs J}'}^{(n)*}\; .
\label{eq:stat2}
\end{equation}
In this equation, $c^{(n)}_{\vec{\bs J}}$ is the $n$-th random sample
for the mode $\vec{\bs J}$.

With $N_{\rm conf}$ samples, the observables we are interested in
(e.g. the inclusive quark spectrum given by eq.~(\ref{eq:Q-LO4})) are
generically approximated by
\begin{equation}
{\cal O}_{N_{\rm conf}}=
{\cal N}\sum_{\vec{\bs f}}
\frac{1}{V^2}\sum_{\vec{\bs J},\vec{\bs J}'}
C_{N_{\rm conf}}(\vec{\bs J},\vec{\bs J}')\;
\big(\psi_{\vec{\bs F}}^{0+}\big|\psi^-_{\vec{\bs J}}\big)
\big(\psi^-_{\vec{\bs J}'}\big|\psi_{\vec{\bs F}}^{0+}\big)\; ,
\end{equation}
where $\vec{\bs F}$ denotes the quantum numbers of the final state,
$\sum_{\vec{\bs f}}$ a partial sum over these quantum numbers (in the
case of eq.~(\ref{eq:Q-LO4}), this partial sum is over the wave number
$\nu$, the spin and color of the produced quark), and ${\cal N}$ a
normalization factor.  $C_{N_{\rm conf}}(\vec{\bs J},\vec{\bs J}')$ is
itself a random number, whose distribution can be determined in the
large $N_{\rm conf}$ limit by a method similar to the derivation of
the central limit theorem. The mean value and variance of this
approximation can be obtained from those of $C_{N_{\rm conf}}(\vec{\bs
  J},\vec{\bs J}')$. Let us first recall that
\begin{equation}
\big<c_{\vec{\bs J}}^{(n)}\big>=0\;,\quad
\big<c_{\vec{\bs J}}^{(n)}c_{\vec{\bs J}'}^{(n')}\big>=0\;,\quad
\big<c_{\vec{\bs J}}^{(n)*}c_{\vec{\bs J}'}^{(n')}\big>=V\,\delta_{nn'}\,\delta_{\vec{\bs J},\vec{\bs J}'}\;.
\label{eq:variance1}
\end{equation}
This leads easily to
\begin{eqnarray}
\big<C_{N_{\rm conf}}(\vec{\bs J},\vec{\bs J}')\big>
&=&V\,\delta_{\vec{\bs J},\vec{\bs J}'}\; ,\nonumber\\
\big<C_{N_{\rm conf}}(\vec{\bs J},\vec{\bs J}')
C_{N_{\rm conf}}(\vec{\bs K},\vec{\bs K}')\big>
&=&\big<C_{N_{\rm conf}}(\vec{\bs J},\vec{\bs J}')\big>
\big<C_{N_{\rm conf}}(\vec{\bs K},\vec{\bs K}')\big>\nonumber\\
&&\qquad+\frac{V^2}{N_{\rm conf}}\,\delta_{\vec{\bs J},\vec{\bs K}'}\;
\delta_{\vec{\bs J}',\vec{\bs K}}\; .
\label{eq:Cstat}
\end{eqnarray}
The formula for the variance is exact if the distribution of
$c_{\vec{\bs J}}$ is Gaussian. Like in the central limit theorem, the
variance of $C_{N_{\rm conf}}$ decreases as $1/N_{\rm conf}$.

From the first of eqs.~(\ref{eq:Cstat}), we obtain the mean value of
${\cal O}_{N_{\rm conf}}$
\begin{equation}
\big<{\cal O}_{N_{\rm conf}}\big>
=
{\cal N}\sum_{\vec{\bs f}}
\frac{1}{V}\sum_{\vec{\bs J}}
\big(\psi_{\vec{\bs F}}^{0+}\big|\psi^-_{\vec{\bs J}}\big)
\big(\psi^-_{\vec{\bs J}}\big|\psi_{\vec{\bs F}}^{0+}\big)\; ,
\end{equation}
which is indeed the exact value of the observable. Using the second of
eqs.~(\ref{eq:Cstat}), we get
\begin{eqnarray}
\big<{\cal O}_{N_{\rm conf}}^2\big>&=&
\big<{\cal O}_{N_{\rm conf}}\big>^2\nonumber\\
&&
+
\frac{1}{N_{\rm conf}}\frac{{\cal N}^2}{V^2}\sum_{\vec{\bs f},\vec{\bs f}'}
\sum_{\vec{\bs J},\vec{\bs K}}
\big(\psi_{\vec{\bs F}}^{0+}\big|\psi^-_{\vec{\bs J}}\big)
\big(\psi^-_{\vec{\bs K}}\big|\psi_{\vec{\bs F}}^{0+}\big)
\big(\psi_{\vec{\bs F}'}^{0+}\big|\psi^-_{\vec{\bs K}}\big)
\big(\psi^-_{\vec{\bs J}}\big|\psi_{\vec{\bs F}'}^{0+}\big)\, .
\nonumber\\
&&
\label{eq:staterr-final}
\end{eqnarray}
We see that the standard deviation of ${\cal O}_{N_{\rm conf}}$
decreases as $1/\sqrt{N_{\rm conf}}$, with a coefficient that has a
non-trivial covariance. It can itself be estimated by the statistical
method as follows~:
\begin{itemize}
\item[{\bf i.}] Define two random linear superpositions of the
  negative energy mode functions~:
  \begin{eqnarray}
    \psi_{1,2}^-\equiv\frac{1}{V}\sum_{\vec{\bs J}}c^{(1,2)}_{\vec{\bs J}}\;\psi_{\vec{\bs J}}^-
  \end{eqnarray}
  with uncorrelated random weights $c_{\vec{\bs J}}^{(1)}$ and $c_{\vec{\bs J}}^{(2)}$,
\item[{\bf ii.}] Evolve these two spinors in time by solving the Dirac equation,
\item[{\bf iii.}] Compute the following quantity~:
  \begin{eqnarray}
    &&
    {\cal N}^2\sum_{\vec{\bs f},\vec{\bs f}'}
    \big(\psi_{\vec{\bs F}}^{0+}\big|\psi^-_1\big)
    \big(\psi^-_2\big|\psi_{\vec{\bs F}}^{0+}\big)
    \big(\psi_{\vec{\bs F}'}^{0+}\big|\psi^-_2\big)
    \big(\psi^-_1\big|\psi_{\vec{\bs F}'}^{0+}\big)=\nonumber\\
    &&\qquad=\Big|{\cal N}\sum_{\vec{\bs f}}
      \big(\psi^-_2\big|\psi_{\vec{\bs F}}^{0+}\big)
      \big(\psi_{\vec{\bs F}}^{0+}\big|\psi^-_1\big)\Big|^2\; ,
      \label{eq:stat4}
  \end{eqnarray}
\item[{\bf iv.}] Repeat the steps {\bf i}-{\bf iii} in order to
  average over the random numbers $c^{(1,2)}_{\vec{\bs J}}$. Since
  this is just an error estimate, a small number of samples is
  sufficient. In practice, one may divide the $N_{\rm conf}$ samples
  already calculated in two subsets, and use these subsets to evaluate
  the error.
\end{itemize}
The standard deviation of ${\cal O}_{N_{\rm conf}}$ is the square root
of the result of this computation, divided by $\sqrt{N_{\rm
    conf}}$. Note that the summand in eq.~(\ref{eq:stat4}) is a
complex number, which can lead to phase cancellations when summing
over the final quantum numbers $\vec{\bs f}$. These cancellations are
more effective for more inclusive observables, thanks to a more
extended sum on $\vec{\bs f}$.

\subsection{Relation to ``low-cost fermions''}
In real-time lattice simulations of fermions, the so-called low-cost
fermion me\-thod \cite{BorsaH1} has been used in several works, e.g.
\cite{AartsS4,AartsS2,SaffiT2,SaffiT1,BergeGS1,KaspeHB1}.  Let us
briefly compare this method with our approach.  In our statistical
method, we compute the following quantity using the stochastic field
\eqref{eq:MC1}:
\begin{equation} \label{ourmethod}
\langle \psi_c^{-\, \dagger} (x) \mathcal{O} \psi_c^- (y) \rangle_c
 = \sum_{\vec{\bs J}} \psi_{\vec{\bs J}}^{-\, \dagger} (x) \mathcal{O} \psi_{\vec{\bs J}}^- (y) \; , 
\end{equation}
where $\vec{\bs J}$ comprises all quantum numbers including momentum,
and $\mathcal{O}$ is a matrix that depends on the observable we wish
to compute.  In the case of the spectrum \eqref{eq:Q-LO4}, ${\cal
  O}\equiv \psi_{\vec{\bs J}'}^{0+} (x) \psi_{\vec{\bs J}'}^{0+\, \dagger}
(y)$.  In the low-cost fermion method, instead of using one stochastic
field \eqref{eq:MC1}, one employs two kinds of stochastic fields
called ``male'' and ``female'' fields:
\begin{equation}
\psi_{_\text{M}} \equiv \frac{1}{\sqrt{2}} \sum_{\vec{\bs J}} 
 \left[ c_{\vec{\bs J}} \psi_{\vec{\bs J}}^+ + d_{\vec{\bs J}} \psi_{\vec{\bs J}}^- \right] , \hspace{10pt} 
\psi_{_\text{F}} \equiv \frac{1}{\sqrt{2}} \sum_{\vec{\bs J}} 
 \left[ c_{\vec{\bs J}} \psi_{\vec{\bs J}}^+ - d_{\vec{\bs J}} \psi_{\vec{\bs J}}^- \right]\; ,
\end{equation}
where $c_{\vec{\bs J}}$ and $d_{\vec{\bs J}}$ are independent random
numbers that have the same variance as eq.~\eqref{eq:variance1}.
Combining these two fields, one can compute
\begin{equation} \label{MFmethod}
-\langle \psi_{_\text{M}}^\dagger (x) \mathcal{O} \psi_{_\text{F}} (y) \rangle
 = \frac{1}{2} \sum_{\vec{\bs J}} \psi_{\vec{\bs J}}^{-\, \dagger} (x) \mathcal{O} \psi_{\vec{\bs J}}^- (y) 
   -\frac{1}{2} \sum_{\vec{\bs J}} \psi_{\vec{\bs J}}^{+\, \dagger} (x) \mathcal{O} \psi_{\vec{\bs J}}^+ (y) \; , 
\end{equation}
instead of eq.~\eqref{ourmethod}.  By using the completeness relation
\begin{equation}
\sum_{\vec{\bs J}}
 \left[ \psi_{\vec{\bs J}}^+ (t,\mathbf{x}) \psi_{\vec{\bs J}}^{+\, \dagger} (t,\mathbf{y})
 +\psi_{\vec{\bs J}}^- (t,\mathbf{x}) \psi_{\vec{\bs J}}^{-\, \dagger} (t,\mathbf{y}) \right]
 = \unit 
\end{equation}
($\unit$ is the unit matrix in the spin, color and position of the
spinors at the time $t$), we can relate the quantities evaluated in
our method \eqref{ourmethod} and in the low-cost fermion method
\eqref{MFmethod} by
\begin{equation}
\langle \psi_c^{-\, \dagger} (x) \mathcal{O} \psi_c^- (y) \rangle
 = -\langle \psi_{_\text{M}}^\dagger (x) \mathcal{O} \psi_{_\text{F}} (y) \rangle 
   +2 \delta (\mathbf{x}-\mathbf{y} ) \; . 
\end{equation}
(for 2 spin states and 2 colors.) Therefore, the two methods provide
the same result\footnote{This is not the case if the initial state is
  not charge neutral.} up to a trivial additive term, that can be
interpreted as a constant vacuum contribution.  However, our method
has two advantages over the low-cost fermion method.  Firstly, it is
numerically less costly than the low-cost fermion method, simply
because it uses only one kind of stochastic field.  Secondly, the
statistical errors are smaller for the evaluation of the spectrum.  In
our method \eqref{ourmethod}, the spectrum is directly obtained from
the statistical ensemble, without any subtraction.  On the other hand,
in the low-cost fermion method \eqref{MFmethod}, one gets directly
access to $\tfrac{1}{2}-{f}$ ($f$ being the fermion occupation
number), and the vacuum 1/2 must be subtracted later.  Because this
vacuum 1/2 also contains statistical errors due to the Monte-Carlo
sampling, the low-cost fermion method suffers from comparatively
larger statistical errors, especially when the value of the occupation
number is small compared to 1/2.

\section{Fermionic mode functions on the light-cone}
\label{sec:mode}
\subsection{Background gauge field}
In order to use the classical-statistical method in the calculation of
the quark spectrum, we should solve the Dirac equation with a
background color field, starting with a free spinor at $x^0=-\infty$.
However, it is not straightforward to do this numerically, due to the
singular nature of the gauge field on the light-cones $x^\pm=0$. The
field strength on these lines is proportional to a $\delta(x^\pm)$,
which cannot be handled easily in a numerical program. Similarly to
the case of gluon production, one should first solve the Dirac
equation analytically up to a surface $\tau=\tau_0\ll Q_s^{-1}$, and
perform the numerical resolution only in the forward light-cone for
$\tau\ge \tau_0$.

\begin{figure}[htbp]
\begin{center}
\hfil
\resizebox*{4cm}{!}{\includegraphics{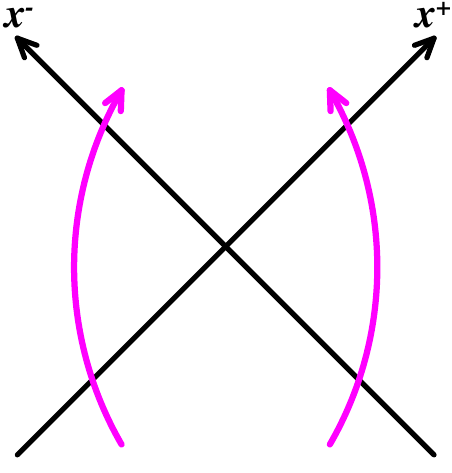}}
\hfil
\hskip 10mm
\resizebox*{4cm}{!}{\includegraphics{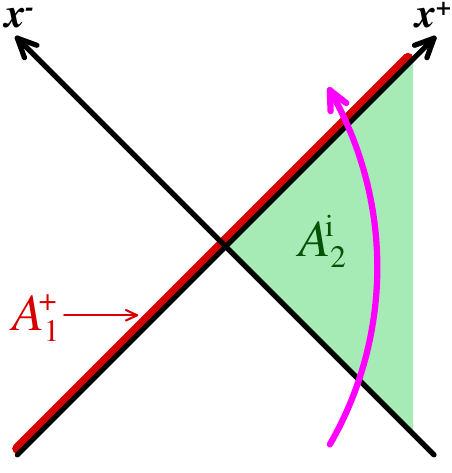}}
\hfil
\end{center}
\caption{\label{fig:bkg}Left~: decomposition of the solution of the
  Dirac equation into left-moving and right-moving partial
  waves. Right~: structure of the gauge field produced by the two
  nuclei in the $A^-=0$ gauge, that we use for solving the Dirac
  equation for the right-moving partial wave. $x^\pm$ denote light-cone
  coordinates, $x^\pm\equiv (t\pm z)/\sqrt{2}$.}
\end{figure}

In this section, we generalize to the case of spinors the derivation
that was performed in ref.~\cite{EpelbG2}. Since the Dirac equation is
linear, its solution can be written as the sum of a left-moving and a
right-moving partial waves, as illustrated in the left diagram in the
figure \ref{fig:bkg}. These two partial waves are totally
independent. We take advantage of this fact to choose the gauge for
the background field differently for each of them, in order to
simplify the resolution of the Dirac equation. Of course, before
adding up the two partial waves in order to construct the full
solution, we must perform a gauge rotation that brings them to a
common gauge.

For the right-moving partial wave, it is convenient to work in the
light-cone gauge $A^-=0$, in the same spirit as what was done in
ref.~\cite{EpelbG2} for the gluons. After having determined the
right-moving spinor at the proper time $\tau_0$ inside the forward
light-cone, we will rotate it to the Fock-Schwinger gauge. The
calculation for the left-moving spinor (to be performed in the
light-cone gauge $A^+=0$) will not be performed here, since the result
can be guessed from the other partial wave by symmetry.  The structure
of the background field in the $A^-=0$ is shown in the right diagram
of the figure \ref{fig:bkg}. It is made of the following two elements~:
\begin{itemize}
\item[{\bf 1.}] the nucleus moving in the $+z$ direction produces a
  field $A_1^+$, proportional to a $\delta(x^-)$ and independent of $x^+$.
\item[{\bf 2.}] in the region $x^+>0, x^-<0$, the nucleus moving in
  the $-z$ direction produces a field $A_1^i$, which has the form of a
  transverse pure gauge,
\begin{equation}
  A_2^i = \frac{i}{g}\,U_2^\dagger\partial^i U_2\; .
  \label{eq:A2}
\end{equation}
\item[{\bf 3.}] in the strip corresponding to the shock-wave of the
  nucleus moving in the $+z$ direction, this field $A_2^i$ receives a color
  precession induced by the first nucleus. This color rotated field reads
\begin{equation}
  \alpha_2^{ia} = \frac{i}{g}U_{1ab}(x^-,\x_\perp)\,(U_2^\dagger\partial^i U_2)_b\; ,
\label{eq:U1A2}
\end{equation}
where
\begin{equation}
U_1(x^-,\x_\perp)
=
{\rm T}\;\exp ig\int_0^{x^-}dz^-\;A_1^+(z^-,\x_\perp)\; .
\label{eq:U1}
\end{equation}
Note that eq.~(\ref{eq:U1A2}) is equivalent to
\begin{equation}
\alpha_2^i\equiv \alpha_2^{ia} t^a = U_1 A_2^i U_1^\dagger\; .
\label{eq:U1A2bis}
\end{equation}
\end{itemize}
Therefore, by starting the evolution at $x^0=-\infty$, the
right-moving partial wave encounters first the field $A_2^i$ and then
the field $A_1^+$. In order to express the quark spectrum, we need the
negative energy spinors, $\psi^-_{\k s a}$ ($\k$ is the 3-momentum of
the incoming quark, $s$ its spin and $a$ its color). Before they
encounter the background color field, they read simply
\begin{equation}
\psi^-_{\k s a}(x)=v_s(\k)\,e^{+ik\cdot x}\; .
\end{equation}
Since the background color field has only a finite jump on the
half-line $x^+=0, x^-<0$, the spinors are continuous across this line,
and the above formulas remain valid up to $x^+=0^+$ (just above this
line).

\subsection{Evolution in the region $x^-<0, x^+>0$}
The next step is to solve the Dirac equation in the region
$x^+>0,x^-<0$. Since the background field is a pure gauge in this
region, the covariant derivative can be written as
\begin{equation}
D^\mu = U_2^\dagger\,\partial^\mu\, U_2\; .
\end{equation}
Therefore, the new spinor defined by $\breve{\psi}_{\k
  s a}^-\equiv U_2\, \psi_{\k s a}^-$ obeys the free Dirac equation~:
\begin{equation}
(i\slpartial-m)\breve{\psi}_{\k s a}^- =0\; .
\end{equation}
The solution of this equation can be expressed in terms of the initial
value of $\breve{\psi}_{\k s a}^-$ on the surface $x^+=0^+$ by the
following Green's formula,
\begin{equation}
\breve{\psi}_{\k s a}^-(x)
=
i
\int_{y^+=0} dy^- d^2\y_\perp\; S_{_R}^0(x,y)\,\gamma^+\,
\breve{\psi}_{\k s a}^-(y)\; ,
\label{eq:green}
\end{equation}
where $S_{_R}^0(x,y)$ is the bare retarded propagator for a quark,
\begin{equation}
S_{_R}^0(x,y)
\equiv
\int \frac{d^4\p}{(2\pi)^4}\;\frac{e^{-ip\cdot(x-y)}}{\slp-m+ip_0\gamma^0\epsilon}\; .
\label{eq:sr}
\end{equation}
Note that, even if we have not restricted the integration range for
the variable $y^-$ in eq.~(\ref{eq:green}), the fact that the support
of the retarded propagator $S_{_R}^0$ is limited to $x^0>y^0,
(x-y)^2\ge 0$ imposes that $y^-\le x^-$. Therefore, for a point $x$
located in the region $x^+>0,x^-<0$ (shaded in green in the figure
\ref{fig:bkg}), only points with $y^-<0$ on the initial surface can
contribute. This is the reason we can solve independently the equation
for the left- and right-moving partial waves.

By introducing the Fourier representation (\ref{eq:sr}) of the
retarded propagator in the Green's formula (\ref{eq:green}), one can
perform most of the integrations analytically except the final
integration over transverse momentum. This leads to
\begin{eqnarray}
{\psi}^-_{\k s a}(x)
&=&
U_2^\dagger(\x_\perp)
\smash{\int\frac{d^2\p_\perp}{(2\pi)^2}}\;e^{i\p_\perp\cdot \x_\perp}\,e^{i(k^+x^-+\frac{M_\p^2}{2k^+}x^+)}\,
\widetilde{U}_2(\p_\perp+\k_\perp)\nonumber\\
&&\qquad\qquad\qquad\qquad
\times\left(1-\gamma^+\frac{p^i\gamma^i+m}{2k^+}\right){\cal P}^+v_s(\k)\; ,
\label{eq:psi-1}
\end{eqnarray}
where $M_\p\equiv\sqrt{\p_\perp^2+m^2}$ is the transverse mass and
where we have introduced the projector ${\cal P}^+\equiv
\gamma^-\gamma^+/2$. We have also introduced the Fourier
transform of the Wilson line $U_2$
\begin{equation}
\widetilde{U}_2(\k_\perp)\equiv \int d^2\y_\perp\; e^{-i\k_\perp\cdot \y_\perp}\; U_2(\y_\perp)\; .
\label{eq:FT-U2}
\end{equation}
One can check that eq.~(\ref{eq:psi-1}) falls back to the free spinor
$v_s(\k)\exp(ik\cdot x)$ if the background field is turned off
($U_2=1$).

\subsection{Evolution across the line $x^-=0~~~ (x^+>0)$}
The next step is to propagate the spinor (\ref{eq:psi-1}) across the
field $A_1^+$ of the nucleus moving in the $+z$ direction. The region
of space-time supporting this field is infinitesimal (since $A_1^+\sim
\delta(x^-)$), but the infinite strength of the gauge field in this
region nevertheless produces a finite change of the spinors. In this
region, there is also an ${\cal O}(1)$ transverse component of the
gauge field, given in eq.~(\ref{eq:U1A2}). The Dirac equation,
\begin{equation}
\left[i(\partial^+-igA_1^+)\gamma^- +i\partial^-\gamma^+-i D^i\gamma^i-m\right]\,\psi^- =0\; ,
\end{equation}
can be separated into a part that depends on the background field
$A_1^+$ and a constraint independent of $A_1^+$ by
multiplying\footnote{One may use the following identities~: ${\cal
    P}^+\gamma^+={\cal P}^-\gamma^-=0$, ${\cal
    P}^-\gamma^+=\gamma^+{\cal P}^+$ and ${\cal
      P}^+\gamma^-=\gamma^-{\cal P}^-$.} it by the projectors ${\cal
    P}^+$ or ${\cal P}^-\equiv \gamma^+\gamma^-/2$~:
\begin{eqnarray}
i\partial^-{\cal P}^+\psi_{\k s a}^-&=&\frac{m-i\gamma^i D^i}{2}\gamma^-{\cal P}^-\psi_{\k s a}^-\nonumber\\
i(\partial^+-igA_1^+){\cal P}^-\psi_{\k s a}^- &=& \frac{\gamma^+(i\gamma^i D^i+m)}{2}\psi_{\k s a}^-\; .
\end{eqnarray}
The first equation, independent of the background field, is a
constraint that relates the two projections of the spinors at every
$x^-$ (note that this equation does not contain the $\partial^+$
derivative). The second equation determines the dynamical evolution
(in the variable $x^-$) of the ${\cal P}^-$ projection under the
influence of the background field $A_1^+$.  Inserting the first
equation into the second gives a second order equation that drives the
evolution of the ${\cal P}^-$ projection,
\begin{equation}
\left(2\partial^-(\partial^+-igA_1^+)-D_\perp^2+m^2\right)\,{\cal P}^-\psi_{\k s a}^-=0\; .
\end{equation}
In this equation, the $\partial^+$ derivative and the field $A_1^+$
are large (inversely proportional to the thickness of the shock-wave
that supports the $A_1^+$ field), while all the other terms do not
have this large factor. Physically, keeping only the term in
$\partial^+-igA_1^+$ leads to the eikonal approximation, where the
fermion would propagate on a straight line along the $x^-$ axis, while
the terms $-D_\perp^2+m^2$ lead to some transverse diffusion with
respect to this axis. In the limit where the thickness of the
shock-wave goes to zero, we can neglect it~:
\begin{eqnarray}
i(\partial^+-igA_1^+){\cal P}^-\psi_{\k s a}^- =0 \; .
\end{eqnarray}
(Note that this approximation is only valid to cross the shock-wave,
and should not be used to evolve at a finite distance from the shock-wave.)
The solution of this equation is very simple,
\begin{equation}
{\cal P}^-\psi_{\k s a}^-(x)=U_1(x^-,\x_\perp)\,{\cal P}^-\psi_{\k s a}^-(0,x^+,\x_\perp)\; ,
\label{eq:-2}
\end{equation}
where the $x^-$-dependent Wilson line $U_1$ was defined in
eq.~(\ref{eq:U1}).  The spinor at $x^-=0$ that appears in the right
hand side is given by eq.~(\ref{eq:psi-1}).
Then, the constraint equation can be solved to give
\begin{equation}
{\cal P}^+\psi_{\k s a}^-(x)
=
-i\int^{x^+}dz^+\;
\frac{m-i\gamma^i D^i}{2}\gamma^-
\,U_1(x^-,\x_\perp)\,{\cal P}^-\psi_{\k s a}^-(0,z^+,\x_\perp)\; .
\label{eq:+2}
\end{equation}
Note that the constraint defines the ${\cal P}^+$ projection of the
spinor only up to an arbitrary function of $x^-$ and $\x_\perp$. This
``integration constant'' can be determined by requesting that we
recover the ${\cal P}^+$ projection of a free spinor when all the
Wilson lines are set to the identity.

\subsection{Transformation to Fock-Schwinger gauge}
In order to gauge transform these spinors to the Fock-Schwinger gauge,
it is sufficient\footnote{In the case of the background gluon field,
  an additional transformation was necessary, see the eq.~(18) of
  ref.~\cite{EpelbG2}. But since the color rotation $\Omega$ that
  characterizes this transformation is equal to the identity for
  proper times $\tau\ll Q_s^{-1}$ (see the eq.~(19) in
  ref.~\cite{EpelbG2}), it has no effect on the spinors.}  to multiply
them by $U_1^\dagger$,
\begin{equation}
\psi_{\k s a\,{}_{\rm FS}}^-(x)
=
U_1^\dagger(\x_\perp)\, \psi_{\k s a}^-(x)\; .
\end{equation}
When this transformation is applied to eqs.~(\ref{eq:-2}) and
(\ref{eq:+2}), the Wilson line $U_1$ appears in two types of
combinations~:
\begin{eqnarray}
U_1^\dagger U_1 = 1\qquad\mbox{and}\qquad
U_1^\dagger D^i U_1\; .
\end{eqnarray}
The second of these structures can be simplified if we recall that
$D^i$ is the covariant derivative built with the field of
eq.~(\ref{eq:U1A2bis}). Therefore,
\begin{eqnarray}
  U_1^\dagger D^i U_1 &=& U_1^\dagger (\partial^i -ig U_1 A_2^i U_1^\dagger) U_1
  \nonumber\\
  &=&\underbrace{\partial^i -ig (A_1^i+A_2^i)}_{D^i_{_{\rm FS}}}\; ,
\end{eqnarray}
where $A_1^i$ is defined in the same way as $A_2^i$ (see the eq.~(\ref{eq:A2})),
\begin{equation}
A_1^i\equiv \frac{i}{g}\,U_1^\dagger\partial^i U_1\; .
\end{equation}
Note that the field $A_1^i+A_2^i$ that appears in this equation is
nothing but the transverse component of the gauge potential in the
Fock-Schwinger gauge at $\tau=0^+$, hence the notation $D_{_{\rm
    FS}}^i$ for the resulting covariant derivative. Therefore, the two
projections of the right-moving negative energy spinors in the
Fock-Schwinger gauge read (at $x^-=0^+$, just above the shock-wave)
\begin{eqnarray}
{\cal P}^-\psi_{\k s a\,{}_{\rm FS}}^-(x)
&=&
U_2^\dagger(\x_\perp)
\smash{\int\frac{d^2\p_\perp}{(2\pi)^2}}\;e^{i\p_\perp\cdot \x_\perp}\,e^{i\frac{M_\p^2}{2k^+}x^+}
\nonumber\\
&&\qquad\qquad\times\;\widetilde{U}_2(\p_\perp+\k_\perp)\,
\frac{p^i\gamma^i-m}{2k^+}\,\gamma^+\, v_s(\k)
\nonumber\\
{\cal P}^+\psi_{\k s a\,{}_{\rm FS}}^-(x)
&=&(i\gamma^iD^i_{_{\rm FS}}-m)\,
U_2^\dagger(\x_\perp)
\smash{\int\frac{d^2\p_\perp}{(2\pi)^2}}\;e^{i\p_\perp\cdot \x_\perp}\,e^{i\frac{M_\p^2}{2k^+}x^+}
\nonumber\\
&&\qquad\qquad\times\;\widetilde{U}_2(\p_\perp+\k_\perp)\,
\gamma^-\,
\frac{p^i\gamma^i-m}{2M_\p^2}\,\gamma^+\, v_s(\k)
\label{eq:psi-FS-right}
\end{eqnarray}
The eqs.~(\ref{eq:psi-FS-right}) for the right-moving spinors must be
completed by a set of similar equations for the left-moving
spinors. These can be obtained from the above formulas by the
following substitutions
\begin{eqnarray}
U_2&\to& U_1\nonumber\\
x^+&\to&x^-\quad,\quad k^+\;\to\; k^-\nonumber\\
{\cal P}^+&\to&{\cal P}^-\quad,\quad {\cal P}^-\;\to\;{\cal P}^+\nonumber\\
\gamma^+&\to&\gamma^-\quad,\quad \gamma^-\;\to\;\gamma^+\; .
\end{eqnarray}
At this point, we have the components of the negative energy mode
functions on the light-cone $\tau=0^+$ (i.e. just after the
collision), which provides all the necessary initial data for studying
their evolution after the collision. As explained in the appendix
\ref{app:IC}, these formulas can also be used as initial conditions at
some initial time $\tau_0>0$, provided that $\tau_0 \ll a_\perp$ where
$a_\perp$ is the transverse lattice spacing used in the numerical
resolution.

\subsection{Mode functions in the $\nu$ basis}
So far, we have derived the mode functions $\psi_{\k s a}$ in terms of
the Cartesian 3-momentum $\k$. However, as explained in the section
\ref{sec:boost},the boost invariance of a high energy collision is
more manifest if one uses the modified spinors defined in
eq.~(\ref{eq:BI-spinors}) and if one further goes to a basis where the
spinors are labeled by the quantum number $\nu$ (Fourier conjugate to
$\eta$) instead of $y$. It is easy to obtain these new mode functions
by the following transformation~:
\begin{equation}
\widehat\psi{}_{\k_\perp\nu s a}\equiv \sqrt{\tau}\;e^{-\frac{\eta}{2}\gamma^0\gamma^3}
\int_{-\infty}^{+\infty}dy\;e^{i\nu y}\; \psi{}_{\k_\perp y s a}\; .
\label{eq:y-to-nu}
\end{equation}
For the right-moving partial waves, the integral that enters in the
transformation of ${\cal P}^\epsilon \psi^-$ is of the form
\begin{equation}
I^{\epsilon}_{_{\rm R}}\equiv\int_{-\infty}^{+\infty}dy\;e^{i\nu y}\;
e^{\epsilon \frac{y}{2}}\;e^{i\alpha e^{-y}}\; ,
\end{equation}
while for the left moving partial waves, one needs
\begin{equation}
I^{\epsilon}_{_{\rm L}}\equiv\int_{-\infty}^{+\infty}dy\;e^{i\nu y}\;
e^{\epsilon \frac{y}{2}}\;e^{i\alpha e^{+y}}\; .
\end{equation}
These integrals can be expressed in terms of the $\Gamma$
function,
\begin{equation}
I_{_{\rm R}}^{\epsilon}
=
(-i\alpha)^{i\nu+\frac{\epsilon}{2}}\,\Gamma\big(-i\nu-\tfrac{\epsilon}{2}\big)
\quad,\quad
I_{_{\rm L}}^{\epsilon}
=
(-i\alpha)^{-i\nu-\frac{\epsilon}{2}}\,\Gamma\big(i\nu+\tfrac{\epsilon}{2}\big)
\; .
\end{equation}

Let us now recapitulate our results for the fermionic mode functions
on the light-cone after this transformation, after summing the
right-moving and left-moving partial waves, and including both the
${\cal P}^+$ and ${\cal P}^-$ projections
\begin{eqnarray}
&&\widehat\psi_{\k_\perp\nu s a\,{}_{\rm FS}}^-(x)
\empile{=}\over{\tau\to 0^+}
-\frac{e^{i\frac{\pi}{4}}}{\sqrt{M_\k}}
\;e^{i\nu\eta}{\int\frac{d^2\p_\perp}{(2\pi)^2}}\;
{\frac{e^{i\p_\perp\cdot \x_\perp}}{M_\p}}\;\nonumber\\
&&\!\!\!\!
\times
\Bigg\{
e^{\frac{\pi\nu}{2}}\Big(\tfrac{M_\p^2\tau}{2M_\k}\Big)^{i\nu}
\!\Gamma(-i\nu\!+\!{\tfrac{1}{2}})
U_2^\dagger(\x_\perp)
\widetilde{U}_2(\p_\perp\!+\!\k_\perp)
\gamma^+
\nonumber\\
&&\!\!\!\!
+
e^{-\frac{\pi\nu}{2}}\Big(\tfrac{M_\p^2\tau}{2M_\k}\Big)^{-i\nu}
\!\Gamma(i\nu\!+\!{\tfrac{1}{2}})
U_1^\dagger(\x_\perp)
\widetilde{U}_1(\p_\perp\!+\!\k_\perp)
\gamma^-\!
\Bigg\}
(p^i\gamma^i\!+\!m)v_s(\k_\perp,y\!=\!0)\,.\nonumber\\
&&
\label{eq:init1}
\end{eqnarray}
In this formula, we have kept only the terms that are non-vanishing in
the limit $\tau\to 0^+$. Therefore, its use should be restricted to
very early times.

\section{Summary and outlook}
\label{sec:summary}
In this paper, we have reconsidered the problem of quark production in
heavy ion collisions in the color glass condensate framework. Our
approach is closely following that of refs.~\cite{GelisKL1,GelisKL2},
which expresses the leading order inclusive quark spectrum in terms of
a set of mode functions of the Dirac equation, but overcomes a number
of limitations of this earlier work.

Firstly, we have defined a basis of fermionic mode functions that are
more appropriate for the boost invariant expanding geometry of a high
energy collision. In particular, since these mode functions are
indexed by the Fourier conjugate $\nu$ to rapidity, they are
especially suitable for a lattice implementation in which one
discretizes the rapidity axis. We have calculated analytically the
value of these mode functions just after the collision, at a proper
time $Q_s\tau\ll 1$ and in the Fock-Schwinger gauge, in terms of the
Wilson lines that represent the classical color background field of
the two colliding nuclei. Thanks to these analytical initial values,
one will not have to deal with crossing the light-cones in the
numerical resolution of the Dirac equation.

Secondly, we have exposed a statistical method for sampling the set of
modes over which one must sum in the calculation of the quark
spectrum. This method ensures that no mode is left out, while
considerably reducing the computing time compared to a complete sum
over all the modes. This approach also provides a more robust way of
estimating the error one makes in the sum over the modes functions.

In a forthcoming paper, we will apply the formalism that we have setup
in the present paper to a study of quark production in two situations.
Firstly, we will present a test of the method in the case of a
background field for which one can solve analytically the Dirac
equation for the mode functions (a constant $SU(2)$ chromo-electrical
field). Then, we will present results on quark production in heavy ion
collisions, in the case where the background color field is given by
the MV model. In order to mitigate the problems caused by the lattice
fermion doubler modes, we will also explain how our framework must be
modified in order to include a Wilson term in the fermionic action.

\section*{Acknowledgements}
FG would like to thank K.~Kajantie and T.~Lappi for numerous
discussions on this problem a long time ago. NT would like to thank
J.~Berges, L.~McLerran, and R.~Venugopalan for  discussions and
comments. NT was partly supported by the Japan Society  for the
Promotion of Science for Young Scientists. FG is supported
by the Agence Nationale de la Recherche project 11-BS04-015-01.

\appendix

\section{Quark spectrum in the Schwinger-Keldysh formalism}
\label{sec:spectrum}
In this appendix, we recall the derivation of the formulas
(\ref{eq:Q-LO0}) for the inclusive quark spectrum, starting from the
standard LSZ reduction formulas. For simplicity, we consider only one
flavor of quarks. Since in the CGC framework, the external sources are
only coupled to gluons and quarks can only produced in quark-antiquark
pairs (the net flavor number is always zero).

\subsection{Single quark pair amplitude}
Firstly, let us consider the amplitude for producing a single
quark-antiquark pair,
\begin{eqnarray}
{\cal M}_1(\p,\q)&\equiv&
\big<Q(\p),\overline{Q}(\q){}_{\rm out}\big|0{}_{\rm in}\big>\nonumber\\
&=&
\big<0{}_{\rm out}\big|b_{\rm out}(\p)d_{\rm out}(\q)\big|0{}_{\rm in}\big>\; .
\label{eq:M1}
\end{eqnarray}
In order to keep the notations concise, we are not writing explicitly
the color and spin indices of the quark and antiquark. The operator
$b_{\rm out}^\dagger(\p)$ (resp. $d_{\rm out}^\dagger(\q)$) creates a
quark of momentum $\p$ (resp. an antiquark of momentum $\q$).

Using the decomposition of the free field operator $\psi_{\rm out}(x)$
as a superposition of free modes, we have
\begin{eqnarray}
b_{\rm out}(\p)&=&
\int d^3\x\; \overline{u}(\p)\gamma^0\;\psi_{\rm out}(t,\x)\;e^{ip\cdot x}
\nonumber\\
d_{\rm out}(\q)&=&
\int d^3\x\; \overline{\psi}_{\rm out}(t,\x)\gamma^0\;v(\q)\;e^{iq\cdot x}
\; .
\label{eq:LSZ}
\end{eqnarray}
(In these formulas, $p^0=\sqrt{\p^2+m^2}$ and $q^0=\sqrt{\q^2+m^2}$.)
Note that the time $t$ at which these formulas are evaluated do not
change the result. Using these formulas, standard manipulations lead
to the LSZ reduction formulas for the single quark pair production
amplitude,
\begin{eqnarray}
{\cal M}_1(\p,\q)&=&
\int d^4x\,d^4y\;
e^{ip\cdot x}\,\overline{u}(\p)(i\stackrel{\rightarrow}{\slpartial}_x-m)
\nonumber\\
&&\times
\big<0{}_{\rm out}\big|{\rm T}\psi(x)\overline{\psi}(y)\big|0{}_{\rm in}\big>\;
(i\stackrel{\leftarrow}{\slpartial}_y+m)\;v(\q)\,e^{iq\cdot y}\; .
\end{eqnarray}
Note that the 2-point correlation function that appears in this
formula is a Feynman (i.e. time-ordered) propagator. Its evaluation to
all orders in the background gluon field cannot be performed in
practice. Indeed, even though it obeys the Dirac equation, its
determination is made extremely complicated by the fact that it must
satisfy mixed boundary conditions, both at $x^0=-\infty$ and at
$x^0=+\infty$.

\subsection{Inclusive quark spectrum}
There is no practical way to calculate the single quark amplitude
considered in the previous subsection, in the presence of a strong
background gluon field, as is the case in the high energy heavy ion
collisions. This is not a big limitation however, because this
quantity is also not very phenomenologically useful in such a
context. Indeed, for light quarks (quark flavors for which $m\lesssim
Q_s$), quark production is not a rare process and more than one quark
pair are produced in a collision. Therefore, the probability $P_1$ of
producing {\sl exactly one} quark pair (given by the square of ${\cal
  M}_1$) is not very interesting. 

Much more useful would be the complete probability distribution, $P_1$,
$P_2$, $P_3$, etc. Unfortunately, calculating them is as complicated
as calculating $P_1$. But it turns out that the moments of the
probability distribution are much easier to compute. Besides the
trivial one ($\sum_{n=0}^\infty P_n=1$), the simplest of these moments
is the first moment, $\sum_{n}nP_n$, that gives the mean number of
produced pairs. A little more information can be gathered by
considering the same quantity in differential form, which is precisely
the quark spectrum. In terms of transition amplitudes, this quantity reads
\begin{eqnarray}
\frac{dN_{\rm q}}{d^3\p}&\equiv&
\frac{1}{(2\pi)^3 2\omega_\p}\,\sum_{n=0}^{+\infty}
(n+1)\frac{1}{(n+1)!^2}\int d\Phi_\q\prod_{i=1}^n d\Phi_{\p_i}d\Phi_{\q_i}
\nonumber\\
&&\!\!\!\times
\left|\big<0{}_{\rm out}\big|b_{\rm out}(\p)d_{\rm out}(\q)b_{\rm out}(\p_1)d_{\rm out}(\q_1)\cdots b_{\rm out}(\p_n)d_{\rm out}(\q_n)\big|0{}_{\rm in}\big>\right|^2\; .
\nonumber\\
&&
\label{eq:spectrum-def}
\end{eqnarray}
where we have used the shorthand
\begin{equation}
d\Phi_\q\equiv \frac{d^3\q}{(2\pi)^3 2\omega_\q}
\end{equation}
for the invariant phase-space of final state quarks and antiquarks.
In this formula, the differential probability for producing $n+1$
quark-antiquark pairs is weighted by the number of quarks ($n+1$), and
then integrated over the phase-space of all the antiquarks and of $n$
of the quarks. This quantity is normalized in such a way that it gives
the mean number of produced quarks after integration over $d^3\p$,
\begin{equation}
\int d^3\p\;\frac{dN_{\rm q}}{d^3\p}=\sum_{n=0}^{+\infty}nP_n
\equiv\left<N_{\rm q}\right>\; .
\end{equation}
Most of eq.~(\ref{eq:spectrum-def}) is in fact the projector on the
subspace of states with net flavor number $-1$,
\begin{eqnarray}
{\bs 1}&\equiv&
\sum_{n=0}^{+\infty}
\frac{1}{n!(n+1)!}\int d\Phi_\q\prod_{i=1}^n d\Phi_{\p_i}d\Phi_{\q_i}\nonumber\\
&&\!\!\!\!
\times\big|[\p_1\cdots\p_n]_{_Q}[\q\q_1\cdots\q_n]_{_{\overline{Q}}}\;{}_{\rm out}\big>\big<[\p_1\cdots\p_n]_{_Q}[\q\q_1\cdots\q_n]_{_{\overline{Q}}}\;{}_{\rm out}\big|\; ,
\end{eqnarray}
that has a trivial action on the state $b_{\rm out}(\p)\big|0{}_{\rm
  in}\big>$ (since this state has also flavor number $-1$), and
eq.~(\ref{eq:spectrum-def}) can then be reduced to the much more
compact form
\begin{equation}
\frac{dN_{\rm q}}{d^3\p}=
\frac{1}{(2\pi)^3 2\omega_\p}\;
\big<0{}_{\rm in}\big| b_{\rm out}^\dagger(\p)b_{\rm out}(\p)\big|0{}_{\rm in}\big>\; .
\label{eq:spectrum-1}
\end{equation}
The interpretation of this formula is quite transparent, since it
amounts to evaluating the expectation value of the final quark number
operator, for a system prepared in the pure state
$\big|0{}_{\rm in}\big>$.

Using eqs.~(\ref{eq:LSZ}), this can be rewritten in terms of the quark
field operator as follows,
\begin{eqnarray}
\frac{dN_{\rm q}}{d^3\p}&=&
\frac{1}{(2\pi)^3 2\omega_\p}\;\int d^4x\,d^4y\;
e^{ip\cdot x}\;\overline{u}(\p)\,(i\stackrel{\rightarrow}{\slpartial}_x-m)
\nonumber\\
&&\times
\big<0{}_{\rm in}\big|\psi(x)\overline{\psi}(y) \big|0{}_{\rm in}\big>\;
(i\stackrel{\leftarrow}{\slpartial}_y-m)\;e^{-ip\cdot y}\,u(\p)\; .
\label{eq:spectrum-2}
\end{eqnarray}
The main differences between this expression and the similar expression
for the amplitude ${\cal M}_1$ are the following~:
\begin{itemize}
\item[{\bf 1.}] The vacuum state is the in-vacuum state on both sides.
\item[{\bf 2.}] The two spinors are not time ordered.
\end{itemize}
These differences in fact lead to considerable simplifications in the
evaluation of this quantity in the presence of a strong background
gluon field. 

The first step is to note that the 2-point correlator that appears in
the integrand of eq.~(\ref{eq:spectrum-2}) is the component
$S_{-+}(x,y)$ of the fermion 2-point function in the Schwinger-Keldysh
formalism. In the presence of a background field $A^\mu$, its tree
level expression (to all orders in the background field) can be
obtained by noticing that it obeys the following equations
\begin{eqnarray}
  &&(i\stackrel{\longrightarrow}{\slD}_x-m)\,{\cal S}_{-+}(x,y)=0\;,\quad
  {\cal S}_{-+}(x,y)\,(i\stackrel{\longleftarrow}{\slD}_y-m)=0\nonumber\\
  &&\lim_{x^0,y^0\to -\infty}{\cal S}_{-+}(x,y)={\cal S}_{-+}^{\rm vacuum}(x,y)\; .
\end{eqnarray}
Next, one should recall the expression of the vacuum propagator ${\cal
  S}_{-+}^{\rm vacuum}(x,y)$,
\begin{equation}
{\cal S}_{-+}^{\rm vacuum}(x,y)
  =
  \sum_{\ontop{s=\uparrow,\downarrow}{a}}\int\frac{d^3\k}{(2\pi)^3 2\omega_\k}\;
  e^{ik\cdot(x-y)}\;
  v_s(\k)\overline{v}_s(\k)\; .
\end{equation}
It is then easy to construct a semi-explicit expression for the
dressed propagator ${\cal S}_{-+}(x,y)$ in terms of a basis of
solutions of the Dirac equation:
\begin{eqnarray}
&&S_{-+}(x,y)=\sum_{\ontop{s=\uparrow,\downarrow}{a}}\int \frac{d^3\k}{(2\pi)^3 2\omega_\k}\;
\psi_{\k s a}(x)\overline{\psi}_{\k s a}(y)
\nonumber\\
&&(i\slD_x-m)\,\psi_{\k s a}(x)=0\quad,\quad\lim_{x^0\to-\infty}\psi_{\k s a}(x)=v_s(\k)\,e^{ik\cdot x}\; .
\end{eqnarray}
When this expression is inserted into eq.~(\ref{eq:spectrum-2}), one
must evaluate the following expression
\begin{equation}
\int d^4x\;e^{ip\cdot x}\;\overline{u}_\sigma(\p)\,
(i\stackrel{\rightarrow}{\slpartial}_x-m)\;\psi_{\k s a}(x)\; .
\label{eq:tmp1}
\end{equation}
It is useful to note that
\begin{equation}
e^{ip\cdot x}\;\overline{u}_\sigma(\p)\,(i\stackrel{\leftarrow}{\slpartial}_x+m)=0\; .
\end{equation}
Adding this identity to the integrand of eq.~(\ref{eq:tmp1}), we
obtain
\begin{equation}
\int d^4x\,e^{ip\cdot x}\,\overline{u}_\sigma(\p)\,
(i\stackrel{\rightarrow}{\slpartial}_x\!-\!m)\,\psi_{\k s a}(x)
=
i\int d^4x\,\partial_\mu\left[
e^{ip\cdot x}\,\overline{u}_\sigma(\p)\gamma^\mu\,\psi_{\k s a}(x)
\right]
\, ,
\label{eq:tmp2}
\end{equation}
which can be rewritten as a 3-dimensional integral since it is the integral
of a total derivative. The boundary at spatial infinity can be dropped
if there are no background fields there. The boundary at $x^0\to
-\infty$ does not contribute because it leads to the vanishing overlap
$u^\dagger(\p)v(\p)$. The only remaining contribution comes from
$x^0\to\infty$,
\begin{equation}
\int d^4x\;e^{ip\cdot x}\;\overline{u}_\sigma(\p)\,
(i\stackrel{\rightarrow}{\slpartial}_x-m)\;\psi_{\k s a}(x)
=
i\lim_{x^0\to+\infty}
\int d^3\x\;
e^{ip\cdot x}\;{u}^\dagger_\sigma(\p)\;\psi_{\k s a}(x)
\; .
\label{eq:tmp3}
\end{equation}
This formula leads immediately to eq.~(\ref{eq:Q-LO0}).

\section{Quark spectrum from Feynman amplitudes}
\label{sec:spectrum-F}
In the previous appendix, we presented a derivation of the quark
spectrum based on fairly standard many-body manipulations: we first
related this spectrum to the expectation value of the quark number
operator, and then we evaluated the latter using the Schwinger-Keldysh
formalism. However, a more elementary derivation is possible, where
the many-body aspects of the problem are treated ``by hand''.  In the
present appendix, we present such an alternate derivation, starting
from the Feynman amplitudes for producing 1,2,3,... quark-antiquark
pairs, and combining them in the appropriate way to obtain the single
quark spectrum. This method is a bit more involved since it requires
to account for all the final state particle permutations, but it has
the advantage of making more tangible the combinatorics that happens
under the hood in the derivation of the appendix \ref{sec:spectrum}.

\subsection{Pair production amplitudes}
The starting point is the amplitude ${\cal M}_1(\p,\q)$ for producing
one quark-antiquark pair, already introduced in
eq.~(\ref{eq:M1}). This amplitude is made of a time ordered 2-point
function connecting the quark of momentum $\p$ and the antiquark of
momentum $\q$, times a disconnected sum of vacuum-vacuum graphs. The
latter is crucial in the presence of a background field, since the sum
of the vacuum-vacuum graphs is not a pure phase (unlike when the
background is the vacuum). In practice, we do not need to calculate
this factor, since it is the same in all amplitudes and can therefore
be determined at the very end by the request that the sum of all
probabilities to produce 0,1,2,3,... quarks is equal to one. For now,
we will simply write
\begin{equation}
{\cal M}_1(\p,\q)\equiv \underbrace{\big<0{}_{\rm out}\big|0{}_{\rm
  in}\big>}_{\mbox{\scriptsize sum of vacuum graphs}}\times\;{\cal M}_1^c(\p,\q)\; ,
\end{equation}
where ${\cal M}_1^c$ is the connected part of the pair production
amplitude (only this factor carries a dependence on the momenta of the
produced quark and antiquark).

In the rest of this appendix, we limit the discussion to the lowest
order for the factor ${\cal M}_1^c$. At this order, it is simply made
of a Feynman propagator connecting the produced quark and antiquark,
dressed by the background field:%
\setbox1\hbox to 2cm{\resizebox*{2cm}{!}{\includegraphics{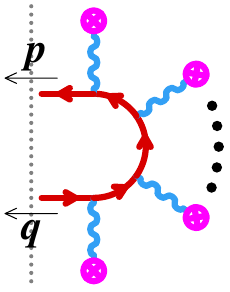}}}%
\begin{equation}
{\cal M}_1^c(\p,\q)=\raise -12mm\box1
\end{equation}
For the sake of simplicity, we will represent this dressed propagator
as follows:%
\setbox1\hbox to 2cm{\resizebox*{2cm}{!}{\includegraphics{M1c}}}%
\setbox2\hbox to 1.4cm{\resizebox*{1.4cm}{!}{\includegraphics{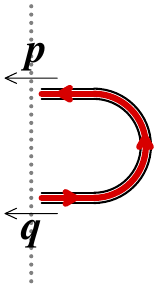}}}%
\begin{equation}
\raise -12mm\box1\quad =\quad \raise -12mm\box2\quad\; .
\end{equation}
A more explicit expression of the connected part of the single pair
production amplitude is
\begin{eqnarray}
  {\cal M}_1^c(\p,\q)=\overline{u}(\p)T_{_F}(p,-q)v(\q)\; ,
  \label{eq:Tf}
\end{eqnarray}
where $T_{_F}$ is the dressed Feynman propagator, amputated of its
final free propagators. In terms of the dressed ($G_{_F}$) and free
($G_{_F}^0$) Feynman propagators, it is given by
\begin{equation}
  G_{_F}=G_{_F}^0+
  G_{_F}^0\;T_{_F}\;G_{_F}^0\; .
\end{equation}
The convention for the momenta in eq.~(\ref{eq:Tf}) is that the
4-momentum $-q$ enters on one side of the propagator and the
4-momentum $p$ exits on the other side.

At the same level of approximation, the amplitude for producing $n$ pairs
is obtained as follows:
\begin{equation}
  {\cal M}_n(\p_1\cdots\p_n,\q_1\cdots\q_n)=
  {\big<0{}_{\rm out}\big|0{}_{\rm
  in}\big>}\sum_{\sigma\in{\mathfrak S}_n}\epsilon(\sigma)\;{\cal M}_1^c(\p_1,\q_{\sigma_1})\cdots
  {\cal M}_1^c(\p_n,\q_{\sigma_n})\; ,
\label{eq:Mn}
\end{equation}
where ${\mathfrak S}_n$ is the symmetry group of the set $[1,n]$. The sum
over all the permutations $\sigma\in{\mathfrak S}_n$ is necessary in order
to account of all the possible ways to connect the quarks with
antiquarks.  $\epsilon(\sigma)$ is the signature of the permutation
$\sigma$, resulting from the signs collected when permuting fermion fields.

\subsection{Final state combinatorics}
It is possible to encapsulate all the information about the
distribution of the produced quarks in the following generating
functional,
\begin{equation}
  {\cal F}[z(\p)]
  \equiv
  \sum_{n=0}^\infty
  \;\int\limits_{{\p_1\cdots\p_n}\atop{\q_1\cdots\q_n}}^+
  \frac{z(\p_1)\cdots z(\p_n)}{n!^2}\;
  \big|{\cal M}_n(\p_1\cdots\p_n,\q_1\cdots\q_n)\big|^2\; ,
\label{eq:Fz}
\end{equation}
where we have used the following shorthand for 1-particle phase-space
integrals:
\begin{equation}
\int\limits_\p^+\equiv\int\frac{d^3\p}{(2\pi)^3 2\omega_\p}=\int\frac{d^4p}{(2\pi)^4}\;2\pi\theta(p^0)\delta(p^2-m^2)\; .
\end{equation}
The $+$ superscript on the integration symbol indicates that we keep
only the positive on-shell energy. Likewise, a $-$ superscript will
indicate that the negative on-shell energy is retained:
\begin{equation}
\int\limits_\p^-\equiv\int\frac{d^4p}{(2\pi)^4}\;2\pi\theta(-p^0)\delta(p^2-m^2)\; .
\end{equation}
From its definition, it is easy to see that the single quark spectrum
is obtained as
\begin{equation}
\frac{dN_{\rm q}}{d^3\p}=\left.\frac{\delta{\cal F}[z]}{\delta z(\p)}\right|_{z\equiv 1}\; .
\end{equation}
Note also that unitary (the sum that all probabilities should be one)
implies that ${\cal F}[z\equiv 1]=1$. Inserting eq.~(\ref{eq:Mn}) into
eq.~(\ref{eq:Fz}), we obtain
\begin{eqnarray}
  {\cal F}[z]
  &=&
 \big|\big<0{}_{\rm out}\big|0{}_{\rm
   in}\big>\big|^2 \sum_n\frac{1}{n!^2}\sum_{\sigma,\sigma'\in{\mathfrak S}_n}
\epsilon(\sigma)\epsilon(\sigma')
 \;
 \int\limits_{{\p_1\cdots\p_n}\atop{\q_1\cdots\q_n}}^+
 z(\p_1)\cdots z(\p_n)\nonumber\\
 &&\times
            {\cal M}_1^{c*}(\p_1,\q_{\sigma_1})\,{\cal M}_1^c(\p_1,\q_{\sigma'_1})\cdots
            {\cal M}_1^{c*}(\p_n,\q_{\sigma_n})\,{\cal M}_1^c(\p_n,\q_{\sigma'_n})\; .
            \label{eq:Fz0}
\end{eqnarray}
When all the spin and Dirac indices are summed over, the product in
the second line forms closed quark loops, from which the spinors can
be eliminated by using $u(\p)\overline{u}(\p)=\slp+m$ and
$v(\q)\overline{v}(\q)=\slq-m$. It is convenient to change $q\to -q$
for all the antiquarks, so that we have:
\begin{eqnarray}
           {\cal F}[z] &=&
 \big|\big<0{}_{\rm out}\big|0{}_{\rm
   in}\big>\big|^2 \sum_n\frac{(-1)^n}{n!}\sum_{\rho\in{\mathfrak S}_n}\epsilon(\rho)
 \int\limits_{{\q_1\cdots\q_n}}^-\prod_{i=1}^nL[z]_{\q_i\q_{\rho_i}}\; ,
 \label{eq:Fz1}
\end{eqnarray}
where we have defined\footnote{We denote $T^*_{_F}(q,p)\equiv\gamma^0\,T_{_F}^\dagger(p,q)\,\gamma^0$.}%
\setbox1\hbox to 2.2cm{\resizebox*{2.2cm}{!}{\includegraphics{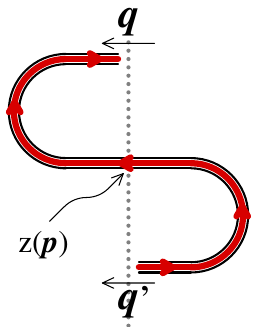}}}%
\begin{equation}
  L[z]_{\q\q'}\equiv\int\limits_\p^+ z(\p)\;
T_{_F}^*(q,p)(\slp+m)T_{_F}(p,q')(\slq'+m)
  \equiv\quad \raise -13mm\box1
  \; .
  \label{eq:Lz}
\end{equation}
In the diagrammatic representation used for this quantity, the dotted
line represents the final state. Right of this line is an amplitude
and left of this line is a complex conjugated amplitude. In order to
go from eq.~(\ref{eq:Fz0}) to eq.~(\ref{eq:Fz1}), we have used the
symmetry of the $n$-quark and $n$-antiquark phase-spaces and the
permutation $\rho$ is defined as $\rho\equiv\sigma^{-1}\sigma'$.

Every permutation $\rho\in{\mathfrak S}_n$ has a unique decomposition
in a product of \emph{cycles}\footnote{We recall the reader that a
  cycle is a circular permutation $1\to \sigma_1 \to
  (\sigma\sigma)_1\to\cdots\to 1$.}. In eq.~(\ref{eq:Fz1}), each of
these cycles will produce a closed quark loop, that depend on $z$
through the quantity $L[z]$ (linear in $z$) defined in
eq.~(\ref{eq:Lz}). The degree in $z$ of such a quark loop is the order
of the cycle (the number of iterations before the cycle returns to the
starting point). For a cycle of order $r$, we will denote the value of
the corresponding quark loop ${\rm tr}\,((L[z])^r)$, where the trace
symbol compactly encapsulates the integrals over all the momenta along
the loop, as well as the contractions over Dirac and color indices.
The important point is that in eq.~(\ref{eq:Fz1}) the product of
$L$'s depends only on the orders of the cycles into which the
permutation $\rho$ can be decomposed: if $\rho=c_1 c_2\cdots c_l$
where the $c_j$'s are cycles of orders $r_1,r_2\cdots r_l$, then we
have
\begin{equation}
  \int\limits_{{\q_1\cdots\q_n}}^-\prod_{i=1}^n L[z]_{\q_i\q_{\rho_i}}
  =
  \prod_{j=1}^l {\rm tr}\,\big((L[z])^{r_j}\big)\; .
\end{equation}
In order to perform the sum over the permutations $\rho$, it is
sufficient to know the number of $\rho$'s that admit a decomposition
into $a_1$ cycles of order 1, $a_2$ cycles of order 2, ..., $a_n$
cycles of order $n$ (with the constraint $a_1+2a_2+\cdots+n a_n=n$),
\begin{equation}
\frac{n!}{a_1!\cdots a_n!}\;\frac{1}{1^{a_1}\cdots n^{a_n}}\; ,
\end{equation}
and its signature
\begin{equation}
  \epsilon(\rho)
  =
  (-1)^n\prod_{j=1}^n (-1)^{a_j}\; .
\end{equation}
Combining these results, we can rewrite\footnote{The final expression
  is reminiscent of the determinant of a Dirac operator, and could
  probably be derived more straightforwardly by path integral
  methods.}
\begin{eqnarray}
&&
  \sum_{n\ge 0}\frac{(-1)^n}{n!}\sum_{\rho\in{\mathfrak S}_n}\epsilon(\rho)
\int\limits_{{\q_1\cdots\q_n}}^-\prod_{i=1}^nL[z]_{\q_i\q_{\rho_i}}
=\nonumber\\
&&\qquad\qquad=\sum_{n\ge 0}\sum_{{a_1+2a_2+\cdots}\atop{\cdots+na_n=n}}
\prod_{j=1}^n\frac{1}{a_j!}\left(-\frac{{\rm tr}\,\big((L[z])^j\big)}{j}\right)^{a_j}
\nonumber\\
&&\qquad\qquad=
\sum_{n\ge 0}\sum_{{a_1+2a_2+\cdots=n}}
\prod_{j=1}^\infty\frac{1}{a_j!}\left(-\frac{{\rm tr}\,\big((L[z])^j\big)}{j}\right)^{a_j}
\nonumber\\
&&\qquad\qquad=
\sum_{p\ge 0}\sum_{{a_1+a_2+\cdots=p}}
\prod_{j=1}^\infty\frac{1}{a_j!}\left(-\frac{{\rm tr}\,\big((L[z])^j\big)}{j}\right)^{a_j}
\nonumber\\
&&\qquad\qquad=
\sum_{p\ge 0}\frac{1}{p!}
\left(-{\rm tr}\,\sum_{j=1}^\infty\frac{(L[z])^j}{j} \right)^p=
\exp \left({\rm tr}\,\ln(1-L[z])\right)\; .
\end{eqnarray}
The second and third lines are equivalent because the constraint
$\sum_{j\ge 1}ja_j=n$ prevents $a_j$'s with $j>n$ from being nonzero.
Going from the third to the fourth line merely corresponds to a
different way of slicing the sum over all $a_j$'s. 

The only missing ingredient is the prefactor $\big|\big<0{}_{\rm
  out}\big|0{}_{\rm in}\big>\big|^2$, that can be trivially determined in
order to satisfy unitarity. The final expression for the generating
functional is therefore
\begin{equation}
  {\cal F}[z]
  =
  \frac{\exp \left({\rm tr}\,\ln(1-L[z])\right)}{\exp \left({\rm tr}\,\ln(1-L[1])\right)}\; .
\end{equation}

Taking a functional derivative leads to the following expression for
the quark spectrum:
\setbox1\hbox to 7.5cm{\resizebox*{7.5cm}{!}{\includegraphics{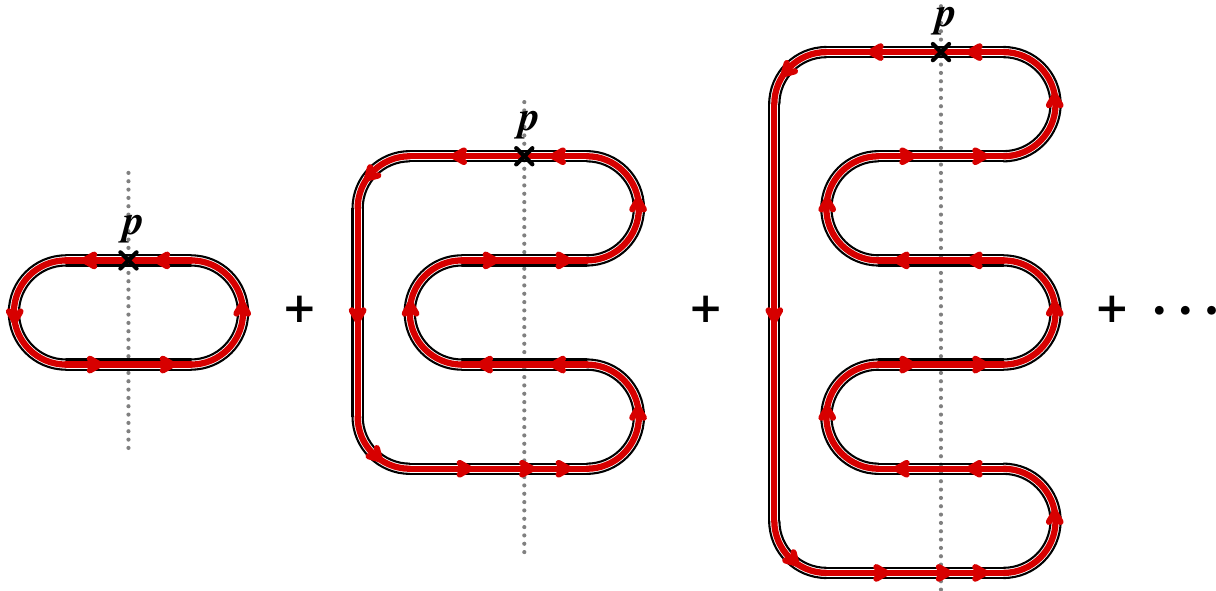}}}%
\begin{eqnarray}
  \frac{dN_{\rm q}}{d^3\p}
  &=&
       -{\rm tr}\,\left(\big(1-L[1]\big)^{-1}\,\left.\frac{\delta L[z]}{\delta z(\p)}\right|_{z\equiv 1}\right)
       \nonumber\\
       &=&
       \raise -16mm\box1
       \label{eq:dNdp}
\end{eqnarray}
At this stage, we have a representation of the quark spectrum in terms
of the object $L[z]$, which is itself quadratic in the \emph{Feynman
  propagator} of a quark in a background field. However, this
expression contains terms of arbitrary order in this propagator,
because of the prefactor $(1-L[1])^{-1}$. Note that in a weak
background field, the quantity $L[1]$ is small and the first term,
quadratic in the Feynman propagator, dominates the sum. This is the
limit where the pair production probability is small and where at most
one pair is produced in a collision. Therefore, the quark spectrum
is almost equal to the differential probability of producing a single
quark, given by the first diagram in the above series. The second, third,
etc... diagrams encode Fermi-Dirac correlations that are only important
when more than one quark are likely to be produced (which is the case
in heavy ion collisions for quarks whose mass is comparable to the
gluon saturation momentum or smaller).

\subsection{Expression in terms of retarded amplitudes}
It turns out that a much simpler expression, quadratic in the
propagator, can be obtained if we rewrite eq.~(\ref{eq:dNdp}) in terms
of the \emph{retarded propagator} of the quark. One can define a
retarded analogue $T_{_R}$ of $T_{_F}$, built with free retarded
propagators instead of free Feynman propagators. The free Feynman and
retarded propagators are related by
\begin{equation}
G_{_F}^0(p)=G_{_R}^0(p)+\underbrace{2\pi(\slp+m)\theta(-p^0)\delta(p^2-m^2)}_{\rho_-(p)}\; .
\end{equation}
If we denote by $V$ one insertion of the background field, the
following two equations define $T_{_F}$ and $T_{_R}$ recursively
\begin{eqnarray}
  T_{_F}&=&V+V G_{_F}^0 T_{_F}=V+ T_{_F}G_{_F}^0V\nonumber\\
  T_{_R}&=&V+V G_{_R}^0 T_{_R}=V+ T_{_R}G_{_R}^0V\; ,
\end{eqnarray}
from which one first obtains:
\begin{eqnarray}
  T_{_F}&=&\big(1-VG_{_F}^0\big)^{-1} V\nonumber\\
  &=& \big(1-VG_{_F}^0\big)^{-1}\big(1-VG_{_R}^0\big) T_{_R}
  =\big(1-VG_{_R}^0-V\rho_-\big)^{-1}\big(1-VG_{_R}^0\big) T_{_R}
  \nonumber\\
  &=&\big(1-(1-VG_{_R}^0)^{-1}V\rho_-\big)^{-1}T_{_R}
  =\big(1-T_{_R}\rho_-\big)^{-1}T_{_R}\; .
\end{eqnarray}
In terms of these compact notations, $L[1]$ can be written as
\begin{eqnarray}
  L[1]
  &=&
  T_{_F}^*\rho_+ T_{_F}\rho_-\; ,
\end{eqnarray}
where $\rho_+(p)\equiv 2\pi(\slp+m)\theta(+p^0)\delta(p^2-m^2)$, and
manipulations similar to above lead to an expression that depends only
on the retarded $T_{_R}$:
\begin{equation}
  L[1]=\big(1-T_{_R}^*\rho_-\big)^{-1}
  \big(T_{_R}^*\rho_+ T_{_R}\rho_-\big)
  \big(1-T_{_R}\rho_-\big)^{-1}\; .
\end{equation}
Note that since $L[z]$ is a linear functional of $z(\p)$, the
derivative $\delta L[z]/\delta z(\p)$ is similar to $L[1]$, but with
the intermediate momentum $p$ fixed instead of being integrated
over. We will denote this as follows:
\begin{equation}
  \frac{\delta L[z]}{\delta z(\p)}=
  \big(1-T_{_R}^*\rho_-\big)^{-1}
  \big(T_{_R}^*\,(\slp+ m)\, T_{_R}\rho_-\big)
  \big(1-T_{_R}\rho_-\big)^{-1}\; .
\end{equation}
In the same fashion, one also obtains the following identities:
\begin{eqnarray}
  \big(1-T_{_R}^*\rho_-\big)\big(1-T_{_R}\rho_-\big)
  &=&1+\big(T_{_R}^*\rho_+ T_{_R}\rho_-\big)\; ,\nonumber\\
  1-L[1] &=& \big(1-T_{_R}^*\rho_-\big)^{-1}\big(1-T_{_R}\rho_-\big)^{-1}\; .
\end{eqnarray}
Therefore,
\begin{equation}
  \frac{dN_{\rm q}}{d^3\p}
  =-{\rm tr}\,\left(\big(1-L[1]\big)^{-1}\;\frac{\delta L[z]}{\delta z(\p)}\right)
  =-{\rm tr}\,\big(T_{_R}^*\,(\slp+ m)\, T_{_R}\rho_-\big)\; .
\end{equation}
The outcome of these algebraic manipulations can be pictorially summarized as%
\setbox1\hbox to 7.5cm{\resizebox*{7.5cm}{!}{\includegraphics{dNdp}}}%
\setbox2\hbox to 1.7cm{\resizebox*{1.7cm}{!}{\includegraphics{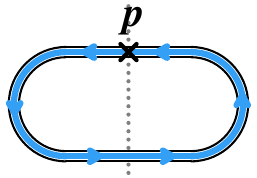}}}%
\begin{equation}
\raise -16mm\box1\quad=\quad\raise -4mm\box2\;\; ,
\end{equation}
where the blue lines in the right hand side are retarded
propagators. The connection with the previous appendix and the formula
(\ref{eq:Q-LO0}) resides in the following identities:
\begin{eqnarray}
  -{\rm tr}\,\big(T_{_R}^*\,(\slp+ m)\, T_{_R}\rho_-\big)
  &=&
  \sum_{\sigma,s}\int_\q^+ \big|\overline{u}_\sigma(\p) T_{_R}(p,-q) v_s(\q)\big|^2\;,
  \nonumber\\
  \overline{u}_\sigma(\p) T_{_R}(p,-q) v_s(\q)&=&
  \lim_{x^0\to+\infty}\int d^3\x\;
  \psi^{0+\dagger}_{\p \sigma}(x^0,\x)\,\psi^-_{\q s}(x^0,\x)\; .
\end{eqnarray}

\section{Evolution from $\tau=0^+$ to $\tau_0>0$}
\label{app:IC}
The eqs.~(\ref{eq:psi-FS-right}) (and their counterparts for the
left-moving partial waves) give the value of the fermionic mode
functions immediately after the collision, one the semi-axis $x^-=0^+,
x^+>0$ for the right-moving partial waves and on the semi-axis
$x^+=0^+, x^->0$ for the left-moving partial waves.  Therefore these
formulas provide for each mode function the complete initial data on
the light-cone $\tau=0^+$.

\begin{figure}[htbp]
\begin{center}
\resizebox*{6cm}{!}{\includegraphics{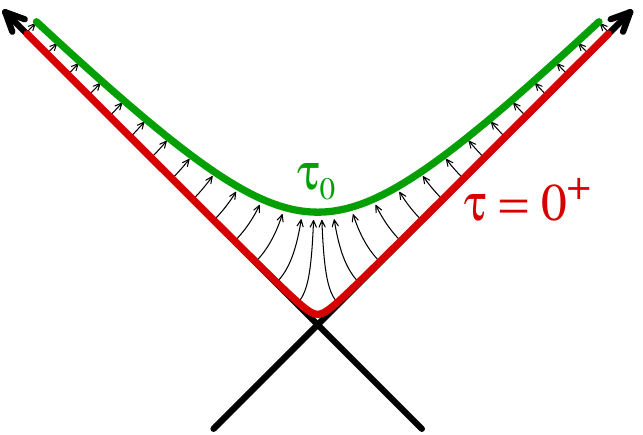}}
\end{center}
\caption{\label{fig:Green-FLC} Evolution between the forward
  light-cone ($\tau=0^+$) and the time $\tau_0$ at which the numerical
  solution of the Dirac equation starts.}
\end{figure} 

However, the transformation to the quantum number $\nu$ (Fourier
conjugate of the spatial rapidity $\eta$) introduces a non-analyticity
in the time dependence, in the form of factors $\tau^{\pm i \nu}$. For
this reason, the numerical resolution of the Dirac equation should be
started at a strictly positive proper time $\tau_0>0$. Therefore, one
should evolve the mode functions from the time $\tau=0^+$ to the time
$\tau_0$ (see the figure \ref{fig:Green-FLC}), by using the Green's
formula for solutions of the Dirac equation,
\begin{equation}
\psi(\tau_0,\eta,\x_\perp)
=
i
\Bigg[
\int\limits_{\ontop{y^-=0^+}{y^+>0}}dy^+
+
\int\limits_{\ontop{y^+=0^+}{y^->0}}dy^-
\Bigg]\int d^2\y_\perp\; 
S_{_R}(\tau_0,\eta,\x_\perp;y)\;\sln\;
\psi(y)\; .
\label{eq:Green-FLC}
\end{equation}
This Green's formula in principle contains the Dirac propagator
dressed by the background color field, which is not known analytically
inside the forward light-cone. However, if one is interested only in
very small propagation times $Q_s\tau_0\ll 1$ one can use the bare
Dirac propagator instead, the effects of the background field on
the evolution of the spinors start becoming important only at
$Q_s\tau_0\gtrsim 1$. But even with a bare propagator, the evaluation
of eq.~(\ref{eq:Green-FLC}) is cumbersome. It turns out that this is
not necessary, if the time $\tau_0$ is chosen small enough.  

For elementary plane waves, the basic formula to justify this is the
following~:
\begin{equation}
\int dy \; e^{i\nu y}\; \left(e^{ik^+ x^-}+e^{ik^-x^+}\right)
\empile{\approx}\over{M_\k\tau_0\ll 1}
\int dy \; e^{i\nu y}\; e^{i(k^+ x^-+ k^-x^+)}
\; ,
\label{eq:Green-approx}
\end{equation}
where $k^\pm=(M_\k/\sqrt{2})\,e^{\pm y}$ and
$x^\pm=(\tau_0/\sqrt{2})\,e^{\pm\eta}$. This formula can be checked by
an explicit calculation of the integrals on both sides, which can be
expressed in terms of Hankel functions for the right hand side and in
terms of Gamma functions in the left hand side, and by doing the
Taylor expansion to first order of the Hankel functions. The
interpretation of this formula is the following:
\begin{itemize}
\item the left hand side is the sum of the left- and right-moving
  partial waves, evaluated as if the time was $\tau=0^+$
  (i.e. neglecting the evolution from $\tau$ to $\tau_0$),
\item the right hand side contains the plane wave evolved to the time
  $\tau_0$ (i.e. the result of using the Green's formula from $\tau$
  to $\tau_0$).
\end{itemize}
In other words, eq.~(\ref{eq:Green-approx}) shows that it is
legitimate to neglect the time evolution of the spinors in the forward
light-cone, provided that the time obeys $M_\k\tau_0\ll 1$. This
exercise shows that if the time $\tau_0$ obeys this condition, it is
sufficient to add up the two partial waves on the light-cone, and to
apply the transformation $y\to \nu$ to their sum.  Let us end this
appendix by an important remark regarding the condition
$M_\k\tau_0\ll 1$: it must be satisfied for all the transverse
masses $M_\k=\sqrt{\k_\perp^2+m^2}$ that can exist in the
problem.  In the lattice discretization of the Dirac equation, this
implies that one must have $\tau_0\ll a_\perp$ where $a_\perp$ is
the transverse lattice spacing.

\section{Conserved inner product}
\label{app:inner}
\subsection{Definition and main properties}
Let us consider a locally space-like surface\footnote{This means that
  if $n^\mu$ is the local orthogonal vector to this surface, then
  $n^\mu n_\mu \ge 0$.} $\Sigma$. For every point $y\in \Sigma$, we
can define an orthogonal vector such that
\begin{eqnarray}
  && n_\mu dy^\mu =0 \mbox{\ \ for any displacement\ }dy^\mu\ \mbox{around\ }y\in\Sigma
\nonumber\\
  && n^0>0
\nonumber\\
&& n_\mu n^\mu=1\; .
\end{eqnarray}
Given two spinors $\psi$ and $\chi$, one can define the following inner
product on $\Sigma$,
\begin{equation}
\big(\psi\big|\chi\big)_{_\Sigma}
\equiv
\int_\Sigma d^3{\bs S}_\y\; \overline{\psi}(y)\,\sln\,\chi(y)\; ,
\label{eq:innerprod}
\end{equation}
where $d^3{\bs S}_\y$ is the 3-dimensional measure\footnote{The
  measure on $\Sigma$ is defined in such a way that $d^3{\bs S}_\y
  d(n\cdot y)$ is the usual 4-dimensional measure $d^4y$. Therefore,
  there is some freedom in how we normalize the vector $n^\mu$,
  provided we change accordingly $d^3{\bs S}_\y$ in such a way that
  $d^3{\bs S}_\y d(n\cdot y)$ is left unchanged.}  on the surface
$\Sigma$.  One sees immediately that this inner product is Hermitean,
\begin{equation}
\big(\psi\big|\chi\big)_{_\Sigma}^*
=
\big(\chi\big|\psi\big)_{_\Sigma}\; .
\end{equation}

The main property of the inner product defined in
eq.~(\ref{eq:innerprod}) is that it is independent of the surface
$\Sigma$ if both $\psi$ and $\chi$ are solutions of the same Dirac
equation\footnote{Note that the quark spectrum involves the inner
  product between a spinor that has evolved over the background field
  and a free spinor. This inner product is therefore not conserved,
  reflecting the fact that the quark yield is time dependent and
  settles to a fixed value only in the limit $\tau\to +\infty$. },
\begin{equation}
(i\slD-m)\psi=0\; .
\end{equation}
(This is true for any real valued background potential $A_\mu$.) From
now on, we can thus drop the subscript $\Sigma$ in our notation for
this inner product. Note also that this inner product is gauge
invariant, since it involves the product of a spinor and the Hermitean
conjugate of another spinor at the same space-time point.

Let us give more explicit expressions of the inner product
\eqref{eq:innerprod} for two important types of surface $\Sigma$. On
a constant $x^0$ surface, it takes the following form,
\begin{equation}
\big(\psi\big|\chi\big)_{{\rm const\ }x^0}
\equiv
\int d^3\y\; {\psi}^\dagger(y)\,\chi(y)\; .
\label{eq:innerprod-x0}
\end{equation}
On a surface of constant proper time, the
integration measure is $d^3{\bs S}_\y\!=\!\tau d\eta d^2\y_\perp$ and the
normal unit vector $n^\mu$ has the following components
\begin{equation}
n^+=\frac{e^{+\eta}}{\sqrt{2}}\quad,\quad n^-=\frac{e^{-\eta}}{\sqrt{2}}\quad,\quad n^i=0\; .
\end{equation}
Therefore, the inner product reads
\begin{equation}
\big(\psi\big|\chi\big)_{\tau}
=
\tau\int d\eta d^2\y_\perp\;
\psi^\dagger(\tau,\eta,\y_\perp)\;e^{-\eta\gamma^0\gamma^3}\;\chi(\tau,\eta,\y_\perp)\; .
\label{eq:innerprod-tau}
\end{equation}

\subsection{Inner product at $x^0=-\infty$}
This conserved inner product can be used as a consistency check for
the various analytic formulas that we have obtained for the mode
functions in the section \ref{sec:mode}. The first step is to evaluate
the inner product on the surface $y^0=-\infty$, where the mode
functions $\psi_{\k s a}^\pm$ are not yet modified by the background
field. We get
\begin{eqnarray}
  \big(\psi_{\k s a}^+\big|\psi_{\k' s' a'}^+\big)&=&(2\pi)^3 2\omega_\k\delta(\k-\k')\delta_{ss'}\delta_{aa'}\; ,\nonumber\\
  \big(\psi_{\k s a}^-\big|\psi_{\k' s' a'}^-\big)&=&(2\pi)^3 2\omega_\k\delta(\k-\k')\delta_{ss'}\delta_{aa'}\; ,\nonumber\\
  \big(\psi_{\k s a}^+\big|\psi_{\k' s' a'}^-\big)&=&0\; .
\label{eq:inner-psi+-}
\end{eqnarray}

\subsection{Inner product at $\tau=0^+$ in LC gauge and $y$ basis}
Now, let us use the eqs.~(\ref{eq:psi-FS-right}) and their
counterparts for the left-moving partial wave in order to check that
the spinors (in light-cone gauge, and in the $y$ basis) evolved to the
forward light-cone are consistent with eqs.~(\ref{eq:inner-psi+-}).
First of all, from the definition of eq.~(\ref{eq:innerprod}), we
immediately see that
\begin{eqnarray}
\big(\psi\big|\chi\big)_{\tau=0^+}
&\equiv&
\sqrt{2}
\int\limits_{y^-=0^+,y^+>0} dy^+d^2\y_\perp\; {\psi}^\dagger(y)\,{\cal P}^-\,\chi(y)
\nonumber\\
&&+\sqrt{2}
\int\limits_{y^+=0^+,y^->0} dy^-d^2\y_\perp\; {\psi}^\dagger(y)\,{\cal P}^+\,\chi(y)
\; .
\label{eq:innerprod-LC}
\end{eqnarray}
In words, on the right branch of the light-cone we need only the
${\cal P}^-$ projection of the spinors, and their ${\cal P}^+$
projection on the left branch of the light-cone.  The other
projections do not contribute to the inner product evaluated on the
light-cone\footnote{Likewise, these projections do not contribute to
  the subsequent evolution of the spinors, because the relevant
  Green's formula also contains a $\sln$.}.

Adding the contributions of the two branches of the light-cone, we
find the following expression for the inner product,
\begin{eqnarray}
\big(\psi_{\k s a}^-\big|\psi_{\k' s' a'}^-\big)_{\tau=0^+}
&=&i\sqrt{2} (2\pi)^2\delta(\k_\perp-\k'_\perp)\delta_{aa'}
\left[
\frac{v^\dagger_s(\k_\perp,y){\cal P}^+v_{s'}(\k_\perp,y')}{k^+-k^{\prime+}+i\epsilon}
\right.\nonumber\\
&&\quad\qquad\qquad\qquad
+
\left.
\frac{v^\dagger_s(\k_\perp,y){\cal P}^-v_{s'}(\k_\perp,y')}{k^--k^{\prime-}+i\epsilon}
\right]\; ,
\end{eqnarray}
where $y,y'$ are the momentum rapidities corresponding to the
3-momenta $\k,\k'$. Using Gordon's identities, one sees that the
imaginary part of the right hand side vanishes. Thanks to
\begin{equation}
\delta(k^+-k^{\prime+})=\frac{\sqrt{2}}{M_\k e^y}\delta(y-y')\quad,\quad
\delta(k^--k^{\prime-})=\frac{\sqrt{2}}{M_\k e^{-y}}\delta(y-y')\;,
\end{equation}
where we denote $M_\k\equiv\sqrt{k_\perp^2+m^2}$, we arrive at
\begin{eqnarray}
\big(\psi_{\k s a}^-\big|\psi_{\k' s' a'}^-\big)_{\tau=0^+}
&=&
(2\pi)^3\delta(y-y')\delta(\k_\perp-\k'_\perp)\delta_{aa'}\frac{1}{M_\k}\nonumber\\
&&\qquad\times\;
{v^\dagger_s(\k_\perp,y)(e^{-y}{\cal P}^++e^y{\cal P}^-)v_{s'}(\k_\perp,y)}\; .
\end{eqnarray}
The second line can be simplified by noticing that
\begin{eqnarray}
\gamma^0\gamma^3 = {\cal P}^+-{\cal P}^-\quad,\quad
\big[{\cal P}^+,{\cal P}^-\big]=0\; .
\end{eqnarray}
From these identities, we obtain easily
\begin{eqnarray}
e^{-y\gamma^0\gamma^3}
=
e^{-y{\cal P}^+}e^{y{\cal P}^-}
=
e^y{\cal P}^-+e^{-y}{\cal P}^+\; .
\end{eqnarray}
Using the fact that $\exp(-y \gamma^0\gamma^3/2)$ acts on spinors as a
boost  of rapidity $-y$ in the $z$ direction, we have
\begin{eqnarray}
v^\dagger_s(\k_\perp,y)\left[e^y{\cal P}^-\!+\!e^{-y}{\cal P}^+\right]
v_{s'}(\k_\perp,y)
&=&
v^\dagger_s(\k_\perp,y)e^{-\frac{y}{2}\gamma^0\gamma^3}\!
e^{-\frac{y}{2}\gamma^0\gamma^3}v_{s'}(\k_\perp,y)\nonumber\\
&=&
v^\dagger_s(\k_\perp,0)v_{s'}(\k_\perp,0)\nonumber\\
&=&2 M_\k \delta_{ss'}\; .
\end{eqnarray}
Therefore, we obtain\footnote{The second equality follows from $\delta(y-y')=\omega_\k\,\delta(k^z-k^{\prime z})$.}
\begin{eqnarray}
\big(\psi_{\k s a}^-\big|\psi_{\k' s' a'}^-\big)_{\tau=0^+}
&=&2 (2\pi)^3 \delta(y-y')\delta(\k_\perp-\k'_\perp)\delta_{ss'}\delta_{aa'}\nonumber\\
&=&2 \omega_\k (2\pi)^3 \delta(\k-\k')\delta_{ss'}\delta_{aa'}\; ,
\end{eqnarray}
which is consistent with the conservation of the inner product. A
similar verification can be performed for the positive energy mode
functions $\psi_{\k s a}^+$.

\subsection{Inner product at $\tau=0^+$ in FS gauge and $\nu$ basis}
\label{app:tau+}
We can perform the same check for the mode functions in the
Fock-Schwinger gauge and $\nu$ basis given by eq.~(\ref{eq:init1}).
Firstly, let us apply the transformation 
\begin{equation}
\psi{}_{\k_\perp\nu s a} \equiv \int dy\;  e^{i\nu y}\; \psi{}_{\k_\perp y s a}
\end{equation}
to eqs.~(\ref{eq:inner-psi+-}) in order to know what to expect for the
inner product in the $\nu$ basis. This trivial calculation tells us that
we should obtain
\begin{equation}
\big(\psi{}_{\k_\perp\nu s a}^-\big|\psi{}_{\k'_\perp \nu' s' a'}^-\big)
=2(2\pi)^4\delta(\nu-\nu')\delta(\k_\perp-\k'_\perp)\delta_{ss'}\delta_{aa'}\; .
\label{eq:inner-FS}
\end{equation}

Let us now calculate this inner product directly from
eq.~(\ref{eq:init1}). The inner product on a surface of constant
$\tau$ is given by eq.~\eqref{eq:innerprod-tau}. It is convenient to
absorb the factor $\tau\exp(-\eta\gamma^0\gamma^3)$ by defining new
spinors that are the original ones boosted to the comoving frame at
the rapidity $\eta$, times a factor $\sqrt{\tau}$,
\begin{equation}
\widehat\psi(\tau,\eta,\y_\perp)\equiv \sqrt{\tau}\,e^{-\frac{\eta}{2}\gamma^0\gamma^3}\;
\psi(\tau,\eta,\y_\perp)\; .
\label{eq:spinor-boost}
\end{equation}
In terms of these boosted spinors, the inner product reads simply~:
\begin{equation}
\big(\psi\big|\chi\big)_{\tau}
=
\int d\eta d^2\y_\perp\;
\widehat\psi^\dagger(\tau,\eta,\y_\perp)
\widehat\chi(\tau,\eta,\y_\perp)\; .
\label{eq:inner-hatpsi}
\end{equation}

When we insert two instances of the formula~(\ref{eq:init1}) in this
equation, we see immediately that the crossed terms vanish because
they contain $(\gamma^+)^2=(\gamma^-)^2=0$. Using the following identities,
\begin{eqnarray}
&&
\int\frac{d^2\p_\perp}{(2\pi)^2}\;\left[\widetilde{U}_2^\dagger(\p_\perp\!+\!\k_\perp)
\widetilde{U}_2(\p_\perp\!+\!\k_\perp')\right]_{aa'}=(2\pi)^2\delta(\k_\perp-\k_\perp')\delta_{aa'}
\nonumber\\
&&
\Gamma(\tfrac{1}{2}+i\nu)\Gamma(\tfrac{1}{2}-i\nu)
=\frac{\pi}{\cosh(\pi\nu)}\; ,
\end{eqnarray}
we first arrive at
\begin{eqnarray}
\big(\psi{}_{\k_\perp\nu s a}^-\big|\psi{}_{\k'_\perp \nu' s' a'}^-\big)
&=&(2\pi)^3\delta(\k_\perp-\k_\perp')\delta(\nu-\nu')\delta_{aa'}
\frac{\pi}{M_\k\cosh(\pi\nu)}
\nonumber\\
&&\!\!\!\!\!\times\,
v_s^\dagger(\k_\perp,y\!=\!0)
\Big[
e^{\pi\nu}\gamma^-\gamma^+
+
e^{-\pi\nu}\gamma^+\gamma^-
\Big]v_{s'}(\k_\perp,y\!=\!0)\; .
\nonumber\\
&&
\label{eq:tmp5}
\end{eqnarray}
The second line can then be simplified as follows~:
\begin{eqnarray}
&&v_s^\dagger(\k_\perp,y\!=\!0)
\Big[
e^{\pi\nu}\gamma^-\gamma^+
+
e^{-\pi\nu}\gamma^+\gamma^-
\Big]v_{s'}(\k_\perp,y\!=\!0)
=
\nonumber\\
&&\qquad=
2\,
v_s^\dagger(\k_\perp,y\!=\!0)
\Big[
e^{\pi\nu}{\cal P}^+
+
e^{-\pi\nu}{\cal P}^-
\Big]v_{s'}(\k_\perp,y\!=\!0)
\nonumber\\
&&\qquad=
2\,
v_s^\dagger(\k_\perp,y\!=\!0)
\,
e^{\pi\nu\gamma^0\gamma^3}
\,v_{s'}(\k_\perp,y\!=\!0)
\nonumber\\
&&\qquad=
2\,
v_s^\dagger(\k_\perp,y\!=\!\pi\nu)
\,v_{s'}(\k_\perp,y\!=\!\pi\nu)
\nonumber\\
&&\qquad=
4\,M_\k\,\cosh(\pi\nu)\,\delta_{ss'}\; .
\label{eq:tmp4}
\end{eqnarray}
Inserting this into eq.~(\ref{eq:tmp5}) gives the expected
result (\ref{eq:inner-FS}).

\section{Abelian case: electron spectrum in QED}
\label{sec:abelian}
The initial value of the mode functions just above the forward
light-cone, given in eq.~(\ref{eq:init1}), can also be used in the
Abelian case.  The analogous problem in QED would be that of the
production of electrons in a high-energy collision of two large
electrical charges $Z_1$ and $Z_2$. When they collide, the
electromagnetic field of the two charges can produce electron-positron
pairs. The spectrum of the produced electrons can be calculated by a
formalism which is very similar to the Color Glass Condensate
considered in this paper. In this description, the two colliding
charges are replaced by the electrical currents they carry along their
trajectories. These currents are the source of electromagnetic fields,
that can be obtained by solving the Maxwell's equations with sources~:
\begin{equation}
\partial_\mu F^{\mu\nu}=J_1^\nu+J_2^\nu\; .
\label{eq:max}
\end{equation}
It is then this electromagnetic field that produces the charged
fermions (in QED, this approach is known as the {\sl equivalent
  photon approximation}.)

Since QED is an Abelian gauge theory, it is much simpler than QCD in
certain respects. Firstly, the Wilson lines $U_1$ and $U_2$ are simply
complex valued phases, that can be commuted at will. Secondly, each of
the colliding charge produces transverse ${\bs E}$ and ${\bs B}$ which
are localized in a shockwave transverse to its trajectory. In the
Fock-Schwinger gauge, the corresponding gauge potential is a pure
gauge in the half-space located after the shockwave. But since
Maxwell's equations are linear, the solution for the 2-charge problem
is the sum of the solutions for individual nuclei, i.e. a sum of two
pure gauge fields, which is itself a pure gauge. In QED, the fields
are thus trivial inside the forward light-cone, unlike the QCD case.

The formula (\ref{eq:init1}), that gives the mode functions just above
the forward light-cone, is therefore sufficient to obtain in closed
form the amplitude. One can evaluate it by computing the inner product
of eq.~(\ref{eq:inner}) at an infinitesimal time $\tau\to 0^+$, since
the evolution at $\tau>0$ is trivial.  The only subtlety when doing so
is that, since there is a non-zero pure gauge field in the forward
light-cone, one should use a gauge rotated free positive energy spinor
instead of $\psi_{\p\sigma}^{0+}$,
\begin{equation}
\psi_{\p\sigma}^{(U_1U_2)+}(x)\equiv
U_1^\dagger(\x_\perp)U_2^\dagger(\x_\perp)\;u_\sigma(\p)\,e^{-ip\cdot x}\; .
\end{equation}
This free spinor must be first transformed as in eq.~(\ref{eq:y-to-nu}),
\begin{equation}
\widehat{\psi}_{\p_\perp\nu\sigma}^{(U_1U_2)+}(x)\equiv\sqrt{\tau}\,
U_1^\dagger(\x_\perp)U_2^\dagger(\x_\perp)
\int dy\; e^{i\nu y}\;e^{-\frac{\eta}{2}\gamma^0\gamma^3}\;u_\sigma(\p)\,e^{-ip\cdot x}\; .
\end{equation}
In order to perform the integral over the momentum rapidity $y$, we
need first to extract the $y$ dependence hidden in the spinor $u_\sigma(\p)$,
\begin{eqnarray}
u_\sigma(\p_\perp,y)=e^{\frac{y}{2}\gamma^0\gamma^3}\,u_\sigma(\p_\perp,0)=
\left[e^{\frac{y}{2}}{\cal P}^++e^{-\frac{y}{2}}{\cal P}^-\right]\,u_\sigma(\p_\perp,0)\; .
\end{eqnarray}
Therefore, we have
\begin{eqnarray}
&&\widehat{\psi}_{\p_\perp\nu\sigma}^{(U_1U_2)+}(x)=\sqrt{\tau}\,
U_1^\dagger(\x_\perp)U_2^\dagger(\x_\perp)
\,e^{i\nu \eta}\,e^{i\p_\perp\cdot\x_\perp}\nonumber\\
&&\qquad\quad\times
\int dy\; e^{i\nu y}\;e^{-iM_\p\tau \cosh(y)}\,\left[e^{\frac{y}{2}}{\cal P}^++e^{-\frac{y}{2}}{\cal P}^-\right]\,u_\sigma(\p_\perp,0)\; ,
\end{eqnarray}
where $M_\p=\sqrt{\p_\perp^2+m^2}$ denotes the transverse mass
(in this equation, we have redefined the integration variable
$y-\eta\to y$). The result of the integration over $y$ can be
expressed in terms of Hankel functions, thanks to
\begin{equation}
\int_{-\infty}^{+\infty}dy\;  e^{-\alpha y}\; e^{-i z \cosh(y)}
=
-i\pi\,e^{-i\frac{\pi\alpha}{2}}\;H^{(2)}_{\alpha}(z)\; ,
\end{equation}
where $H_\alpha^{(2)}(z)\equiv J_\alpha(z)-iY_\alpha(z)$. In the limit
$\tau\to 0^+$, we need only the beginning of the Taylor expansion of
the Hankel function,
\begin{equation}
H^{(2)}_\alpha(z)\empile{=}\over{z\to 0^+}
\frac{i}{\sin(\pi\alpha)}\left[
\left(\frac{z}{2}\right)^{-\alpha}\frac{1}{\Gamma(1-\alpha)}
-
\left(\frac{z}{2}\right)^\alpha\frac{e^{i\pi\alpha}}{\Gamma(1+\alpha)}
\right]\; .
\end{equation}
Note that when $\alpha$ has a non-zero real part, only one of the two
terms dominates when $z\to 0$, depending on the sign of this real
part.  Therefore, in the combination
$\sqrt{z}H^{(2)}_{-i\nu-\frac{\epsilon}{2}}(z)$, only
one of the two terms survives when $z\to 0^+$,
\begin{eqnarray}
&&\sqrt{z}\;e^{-i\frac{\pi}{2}(-i\nu-\frac{1}{2})}\;H^{(2)}_{-i\nu-\frac{1}{2}}(z)
\empile{=}\over{z\to 0^+}
\frac{\sqrt{2}\,e^{i\frac{\pi}{4}}\,e^{\frac{\pi\nu}{2}}}{\cosh(\pi\nu)\Gamma(\frac{1}{2}-i\nu)}\;
\left(\frac{z}{2}\right)^{-i\nu}
\nonumber\\
&&\sqrt{z}\;e^{-i\frac{\pi}{2}(-i\nu+\frac{1}{2})}\;H^{(2)}_{-i\nu+\frac{1}{2}}(z)
\empile{=}\over{z\to 0^+}
\frac{\sqrt{2}\,e^{i\frac{\pi}{4}}\,e^{-\frac{\pi\nu}{2}}}{\cosh(\pi\nu)\Gamma(\frac{1}{2}+i\nu)}\;
\left(\frac{z}{2}\right)^{i\nu}\; .
\end{eqnarray}
Therefore,
\begin{eqnarray}
&&\widehat{\psi}_{\p_\perp\nu\sigma}^{(U_1U_2)+}(x)
\empile{=}\over{\tau\to 0^+}
\pi e^{-i\frac{\pi}{4}}\,U_1^\dagger(\x_\perp)U_2^\dagger(\x_\perp)\;
\sqrt{\frac{2}{M_\p}}\,
\frac{e^{i\nu \eta}\,e^{i\p_\perp\cdot\x_\perp}}{\cosh(\pi\nu)}
\nonumber\\
&&\quad\times
\left[
\frac{e^{\frac{\pi\nu}{2}}}{\Gamma(\tfrac{1}{2}-i\nu)}
\left(\tfrac{M_\p\tau}{2}\right)^{-i\nu}{\cal P}^+
+
\frac{e^{-\frac{\pi\nu}{2}}}{\Gamma(\tfrac{1}{2}+i\nu)}
\left(\tfrac{M_\p\tau}{2}\right)^{i\nu}{\cal P}^-
\right]
\,u_\sigma(\p_\perp,0)\; .
\nonumber\\
&&
\end{eqnarray}
Similarly, the negative energy spinors evolved from the remote past up
to $\tau=0^+$ read,
\begin{eqnarray}
&&\widehat\psi_{\k_\perp\mu s}^-(x)
=
-\frac{e^{i\frac{\pi}{4}}}{\sqrt{M_\k}}
\;e^{i\mu\eta}{\int\frac{d^2\q_\perp}{(2\pi)^2}}\;
{\frac{1}{M_\q}}\;\nonumber\\
&&\quad\times
\Bigg\{
\left(\frac{M_\q^2\tau}{2M_\k}\right)^{i\mu}
e^{\frac{\pi\mu}{2}}\,
\Gamma(\tfrac{1}{2}-i\mu)\,
U_2^\dagger(\x_\perp)
\widetilde{U}_2(\q_\perp+\k_\perp)\,
\gamma^+
\nonumber\\
&&
\qquad+
\left(\frac{M_\q^2\tau}{2M_\k}\right)^{-i\mu}
e^{-\frac{\pi\mu}{2}}\,
\Gamma(\tfrac{1}{2}+i\mu)\,
U_1^\dagger(\x_\perp)
\widetilde{U}_1(\q_\perp+\k_\perp)\,
\,\gamma^-\, 
\Bigg\}\nonumber\\
&&\qquad\qquad
\times e^{i\q_\perp\cdot \x_\perp}\,(q^i\gamma^i+m)\,v_s(\k_\perp,y=0)\; .
\label{eq:init2}
\end{eqnarray}

Using ${\cal P}^+\gamma^+={\cal P}^-\gamma^-=0$, the inner product
between ${\psi}_{\p_\perp\nu\sigma}^{(U_1U_2)+}$ and
$\psi_{\k_\perp\mu s}^-$ reads
\begin{eqnarray}
&&
\!\!
\big({\psi}_{\p_\perp\nu\sigma}^{(U_1U_2)+}\big|\psi_{\k_\perp\mu s}^-\big)
=
2\pi\delta(\nu-\mu)\frac{\pi\,e^{i\frac{\pi}{2}}}{\cosh(\pi\nu)}
\sqrt{\tfrac{2}{M_\p M_\k}}
\int\frac{d^2\q_\perp}{(2\pi)^2}\,\frac{1}{M_\q}
\,
u_\sigma^\dagger(\p_\perp,0)\nonumber\\
&&\qquad\times\Big\{
\left(\tfrac{M_\k M_\p}{M_\q^2}\right)^{i\nu}
\widetilde{U}_2(\p_\perp\!-\!\q_\perp)\widetilde{U}_1(\q_\perp\!+\!\k_\perp)\,\gamma^-
\nonumber\\
&&\qquad\;
+
\left(\tfrac{M_\k M_\p}{M_\q^2}\right)^{-i\nu}
\widetilde{U}_1(\p_\perp\!-\!\q_\perp)\widetilde{U}_2(\q_\perp\!+\!\k_\perp)\,\gamma^+
\Big\}\,(q^i\gamma^i+m)\,v_s(\k_\perp,0)\; .
\nonumber\\
&&
\end{eqnarray}
In order to compare with existing results in the literature
(e.g. eq.~(52) in ref.~\cite{BaltzM1}), one should perform the inverse
transformation $\nu\to y_p,\mu\to y_k$ to return to momentum rapidity
variables~:
\begin{eqnarray}
\big({\psi}_{\p\sigma}^{(U_1U_2)+}\big|\psi_{\k s}^-\big)
=
\int\frac{d\nu d\mu}{(2\pi)^2}
\;
e^{i\nu y_p}e^{-i\mu y_k}\;
\big({\psi}_{\p_\perp\nu\sigma}^{(U_1U_2)+}\big|\psi_{\k_\perp\mu s}^-\big)
\; .
\end{eqnarray}
Thanks to the following formula,
\begin{equation}
\int\frac{d\nu}{2\pi}\;\frac{e^{i\nu z}}{\cosh(\pi\nu)}=
\frac{1}{2\pi\,\cosh(\frac{z}{2})}\; ,
\end{equation}
it is easy to perform these integrals and one obtains
\begin{eqnarray}
&&\big({\psi}_{\p\sigma}^{(U_1U_2)+}\big|\psi_{\k s}^-\big)
=i\sqrt{2}
\int\frac{d^2\q_\perp}{(2\pi)^2}\;u^\dagger_\sigma(\p_\perp,y=0)\nonumber\\
&&\quad\times\,\Bigg\{
\frac
{\widetilde{U}_2(\p_\perp-\q_\perp)\widetilde{U}_1(\q_\perp+\k_\perp)e^{\frac{y_p-y_k}{2}}\gamma^-}
{M_\q^2+2k^-p^+}
\nonumber\\
&&\quad\;\;
+
\frac
{\widetilde{U}_1(\p_\perp-\q_\perp)\widetilde{U}_2(\q_\perp+\k_\perp)e^{\frac{y_k-y_p}{2}}\gamma^+}
{M_\q^2+2k^+p^-}
\Bigg\}(m+q^i\gamma^i)v_s(\k_\perp,y=0)\; .
\nonumber\\
&&
\end{eqnarray}
The final step to compare with ref.~\cite{BaltzM1} is to use the identities
\begin{eqnarray}
&&
e^{\frac{y_p-y_k}{2}}u^\dagger_\sigma(\p_\perp,y=0)\gamma^-(m+q^i\gamma^i)v_s(\k_\perp,y=0)
\nonumber\\
&&\qquad\qquad\qquad
=u^\dagger_\sigma(\p_\perp,y_p)\gamma^-(m+q^i\gamma^i)v_s(\k_\perp,y_k)
\nonumber\\
&&
e^{\frac{y_k-y_p}{2}}u^\dagger_\sigma(\p_\perp,y=0)\gamma^+(m+q^i\gamma^i)v_s(\k_\perp,y=0)
\nonumber\\
&&\qquad\qquad\qquad
=u^\dagger_\sigma(\p_\perp,y_p)\gamma^+(m+q^i\gamma^i)v_s(\k_\perp,y_k)\; ,
\end{eqnarray}
thanks to which we finally obtain
\begin{eqnarray}
&&\big({\psi}_{\p\sigma}^{(U_1U_2)+}\big|\psi_{\k s}^-\big)
=i\sqrt{2}
\int\frac{d^2\q_\perp}{(2\pi)^2}\;u^\dagger_\sigma(\p_\perp,y_p)\nonumber\\
&&\qquad\times\,\Bigg\{
\frac
{\widetilde{U}_2(\p_\perp-\q_\perp)\widetilde{U}_1(\q_\perp+\k_\perp)\gamma^-}
{M_\q^2+2k^-p^+}
\nonumber\\
&&\qquad\quad\;
+
\frac
{\widetilde{U}_1(\p_\perp-\q_\perp)\widetilde{U}_2(\q_\perp+\k_\perp)\gamma^+}
{M_\q^2+2k^+p^-}
\Bigg\}(m+q^i\gamma^i)v_s(\k_\perp,y_k)\; .
\nonumber\\
&&
\end{eqnarray}
This formula is identical to the eq.~(52) in ref.~\cite{BaltzM1}. Note
that in order to recover this known result, it was crucial to gauge
rotate the free spinor used in the projection, because the gauge field
inside the forward light-cone is a non-zero pure gauge in the
Fock-Schwinger gauge.


\end{document}